\documentclass{aa}
\bibpunct{(}{)}{;}{a}{}{,}
\usepackage[varg]{txfonts}
\usepackage{graphicx}  
\usepackage[dvipsnames]{xcolor}    
\usepackage{multirow}
\usepackage{placeins}

\newcommand{\Msun}[0]{\rm{M_\odot}}
\newcommand{\Meart}[0]{\rm{M_\oplus}}
\newcommand{\Lsun}[0]{\rm{L_\odot}}

\newcommand{\FeH}[0]{\rm{[Fe/H]}}

\newcommand{\Teff}[0]{T_\mathrm{eff}}

\begin{document}

\title{Chemical paradox in a binary system: Exploring metal enrichment in  HD~81809B}

\author{Nuno Moedas\inst{1}\thanks{nammo@dtu.dk}, \and Maria Pia Di Mauro\inst{2}}

\institute{DTU Space, Technical University of Denmark, Elektrovej 327, Kgs. Lyngby, 2800, Denmark \and INAF-IAPS, Via del Fosso del Cavaliere 100, 00133 Roma, Italy}

\date{Received XXX / Accepted YYY}

\abstract 
{The HD~81809 system presents peculiar chemical composition
with a large metallicity difference between its two components: the primary has low metallicity ($\mathrm{[Fe/H]} = -0.57$~dex), while the secondary has approximately solar metallicity ($\mathrm{[Fe/H]} = 0.0$~dex).}
{This study investigates whether the chemical enrichment of HD~81809B can be reconciled by a planetary engulfment event,
consistent with the star’s evolutionary constraints.}
{Using Modules for Experiments in Stellar Astrophysics (MESA) code, we model HD~81809B introducing  accretion events with different ranges of masses and chemical mixtures to understand their impact in the surface chemical composition.}
{Our simulation shows that reproducing the observed surface $\FeH$ in HD~81809B requires a substantial accretion event. To reach the observed metallicity, the star must engulf 25 to 75 $\Meart$ of metals near its current age. Accretion brings the effective temperature of stellar models closer to the observed one; however, the lithium results over-enriched. In order to reproduce the abundance of this element, the star must accrete less than 6 $\Meart$. }
{These results suggest that a planetary accretion event is plausible in the evolution of HD~81809B; 
  however, the chemical composition of the accreted material may differ from the assumptions adopted here. Nevertheless, this study emphasizes the importance of considering such external events when modeling chemically nomalous binary systems.}
\keywords{Stars: abundances -- Stars: evolution -- Accretion, accretion disks -- Planet-star interactions -- Binaries: general -- Stars: solar-type}

\titlerunning{Chemical paradox in a binary system}
\authorrunning{Nuno Moedas \& Maria Pia Di Mauro}
\maketitle

\section{Introduction}
\label{sec:Intro}
Binary systems provide fundamental constraints for stellar evolution because their components formed simultaneously from the same molecular cloud and therefore share the same age and initial chemical composition. Any observed differences between the components can thus reveal the effects of stellar mass, internal transport processes, or external events such as mass transfer or planetary engulfment. When additional constraints such as asteroseismic measurements and dynamical masses are available, binary systems become powerful laboratories for testing stellar structure and evolution models

HD~81809 is a particularly valuable binary system consisting of two G-type stars at different stages of evolution. The primary component (HD~81809A) is a subgiant (SG) star, while the secondary (HD~81809B) is a main-sequence (MS) star. The system has attracted significant attention because the primary exhibits a well-defined magnetic activity cycle \citep[e.g.][]{Orlando2017,Egeland2018,Mauro2026}, making it an important target for studying stellar dynamos in solar-type stars. Early determinations of the stellar masses based on radial velocities were presented by \cite{Duquennoy1988}, who estimated values of $1.7 \pm 0.64~\Msun$ and $1.0 \pm 0.25~\Msun$ for the primary and secondary, respectively. 

Using stellar evolutionary models, \cite{Fuhrmann2018} derived a mass of $1.39\pm0.09~\Msun$ and an age of $3.2$~Gyr for the primary component. However, this relatively young age appears inconsistent with the long observed rotation period ($40.2 \pm2.3$~days; \citealt{Donahue1993,Donahue1996,Egeland2018}), as magnetic braking would not have had sufficient time to slow the stellar rotation to the measured value. 
In addition, HD~81809A exhibits an enhancement in $\alpha$ elements ($\rm[Fe/Mg] = -0.35\pm0.05~dex$) that is not expected from standard Galactic chemical evolution trends. To reconcile these discrepancies, \cite{Fuhrmann2018} proposed that the primary may have experienced a merger event involving a subgiant and a lower-mass MS star.

Recently, \cite{Mauro2026} (hereafter Paper~I) revisited the system using updated spectroscopic, photometric, and radial velocity data, deriving dynamical masses of $0.92 \pm 0.09~\Msun$ and $0.79 \pm 0.06~\Msun$ for components A and B, respectively.
 In addition, using observations from the Transiting Exoplanet Survey Satellite (TESS), they detected solar-like oscillations in both components and measured the large frequency separation, $\Delta \nu$, and the frequency of maximum oscillation power, $\nu_\mathrm{max}$. By modeling these seismic constraints with grids of stellar models, they inferred an age of approximately 10~Gyr for the primary, significantly older than previous estimates and more consistent with Galactic chemical evolution expectations.

Despite this progress, several inconsistencies remain. In particular, Paper~I was unable to simultaneously reproduce the observed luminosity of the primary ($5.10\pm0.14~\Lsun$), obtaining instead a lower modeled value of $\sim3.3~\Lsun$. The authors noted that including the observed luminosity as a constraint may lead to an overestimation of the stellar mass by $\sim0.3~\Msun$ and of the effective temperature by about 300~K. 
From the spectroscopic analyses in Paper~I, they identified this star as being metal-poor ($\FeH=-0.57\pm0.18$~dex), and enriched in alpha elements compared to the Sun ($\rm [Mg/Fe]=0.38\pm0.24$~dex). They suggested adopting a different chemical mixture and appropriate opacity tables, as these have a significant impact on the inferred mass and age of HD~81809A.

The secondary component presents an additional challenge. Paper~I reported a marked difference in the surface chemical composition of the two stars: the primary shows a metallicity of $\FeH=-0.57\pm0.18$~dex, while the secondary is compatible with solar metallicity ($\FeH=0.00\pm0.11$~dex). Such a large chemical difference is not expected in a binary system formed from the same molecular cloud \citep{Adibekyan2018}. Moreover, chemical transport mechanisms such as atomic diffusion and other mixing processes are not predicted to produce discrepancies of this magnitude \citep{Deal2015,Ramirez2019,Liu2021,moedas2022}. 

To explain this dichotomy, Paper~I explored two formation scenarios for HD~81809B. In the first scenario, the secondary is assumed to be a captured star formed in a different molecular cloud. This solution yields an age of $15.77\pm4.11$~Gyr, formally compatible within uncertainties but exceeding the age of the Universe at the nominal value. In addition, the inferred mass of the secondary becomes inconsistent with dynamical constraints. In the second scenario, both stars are assumed to have formed together. This yields a mass of $\sim0.83~\Msun$ for the secondary and an age closer to that of the primary, but fails to reproduce the observed surface metallicity and overestimates the effective temperature. 

Previous investigations into chemical inhomogeneities within stellar binaries have yielded conflicting results. While some observational studies report distinct abundance differences—particularly in planet-hosting systems \citep[e.g.,][]{Saffe2017,Oh2018,Ramirez2019,Nagar2020}—others find no significant discrepancies \citep[e.g.,][]{LiuF2014,Saffe2015,Mack2016}. Theoretically, minor variations on the order of $0.1-0.2$~dex can be naturally explained by atomic diffusion competing with other internal transport processes \citep{Deal2015,Ramirez2019,Liu2021}. However, the $\sim0.57$~dex discrepancy observed in the HD~81809 system is simply too extreme for these standard mechanisms. Such a large difference can only be theoretically reproduced in F-type stars if atomic diffusion is treated in isolation \citep{moedas2022}, making it an unlikely product of standard stellar evolution. 

Driven by the detection of a debris disk in the system, Paper~I hypothesized that a recent accretion event could have enriched the surface layers of HD~81809B.
This hypothesis is further supported by the secondary's high surface lithium abundance ($\rm[Li/H]=0.69\pm0.18$~dex). Standard evolution models predict significant Li depletion in 
$0.7-0.9~\Msun$ stars \citep{Cummings2017,Sun2023}. Although stars 
in this mass range may preserve lithium enrichment signatures for a few Gyr \citep{Sevilla2022}, detecting the consequences of an accretion event is generally rare due to the combined effects of convection, thermohaline mixing, and atomic diffusion \citep{Theado2009,Vauclair2012,Deal2015, Behmard2022,
Soares2025}.

The HD~81809 system, with its extreme chemical peculiarities, thus provides a unique laboratory to test planetary engulfment in stellar models.
In this work, we explore whether an accretion event involving metal-rich material can account for the observed surface enrichment of HD~81809B. We employ stellar evolution models including detailed chemical transport processes to test whether engulfment scenarios can simultaneously reproduce the observed metallicity and lithium abundance of the secondary, while remaining consistent with the seismic and dynamical constraints of the system.

This study is structured as follows: In Sect.~\ref{sec:Obsevable} we present the observed parameters of the HD~81809 system and previous studies. The stellar physics and accretion mechanism adopted are explained in Sect.~\ref{sec:Stellar_models}. In Sect.~\ref{sec:accre_models} we present the results of including accretion in HD~81809B models. We discuss other possible scenarios for chemical discrepancy in Sect.~\ref{sec:discussion}. We conclude in Sect.~\ref{sec:conclusion}. 

\section{HD81809 system}
\label{sec:Obsevable}

Paper~I provided the most recent comprehensive analysis of the HD~81809 system. They derived fundamental stellar parameters for both components from combined asteroseismic, spectroscopic, photometric, radial velocity, and astrometric data (summarized in Table~\ref{tab:params}). 
Using these updated observational constraints, Paper~I inferred the fundamental properties of both stars employing stellar evolutionary models from the \textsc{CLES} code \citep{Scuflaire2008} and \textsc{MESA} \citep{Paxton2011,Paxton2013,Paxton2015,Paxton2018,Paxton2019}. In the present work, we adopt the MESA models from Paper~I as a reference to investigate the impact of accretion on the surface composition of HD~81809B and assess whether the observed chemical variations can be reproduced. The inferred stellar properties, including both scenarios for HD~81809B, are listed in Table~\ref{tab:star_inf}.

Our analysis focuses on the secondary star using model B2 from Paper~I (hereafter denoted as M2), which was identified as the more physically plausible scenario, although does not well reproduce the observed surface $\FeH$ and $\Teff$ as model B1 from  Paper~I (hereafter denoted as M1).

Using model M2 as a baseline, we introduce mass accretion in the evolution of HD~81809B to examine whether this process can improve the agreement with the observed surface parameters.

\begin{table}[]
\caption{Stellar parameters from observation provided in Paper I.}
\centering
\label{tab:params}
\begin{tabular}{lrr}
\hline
Parameter & Star A & Star B \\
\hline
$M~(\rm M_{\odot})$    & $ 0.92 \pm 0.09  $ & $ 0.79 \pm 0.06  $ \\
$\log g$ (dex)          & $ 3.77 \pm 0.05  $ & $ 4.22 \pm 0.03  $ \\
$T_{\rm eff}$ (K) & $ 5580 \pm 140   $ & $ 5520 \pm 150   $ \\
$L~(\rm L_{\odot})$     & $ 5.10 \pm 0.14  $ & $ 0.92 \pm 0.05  $ \\
$\rm \Delta\nu\ (\mu Hz)$    & $ 43.32 \pm 3.91  $ & $ 97.75 \pm 4.49  $ \\
$\rm \nu_{max}\ (\mu Hz)$    & $ 708.74^{3.23}_{-3.74}$ & $ 2098.07^{+3.07}_{2.83}$ \\
$\FeH$ (dex)   & $-0.57\pm0.18$ & $0.00\pm0.11$ \\
$[\rm Li/H]$ (dex)   & $-0.05\pm0.18$ & $0.69\pm0.11$ \\
\hline
\end{tabular}
\end{table}

\begin{table*}[t!]
\centering
\caption{Stellar properties inferred in Paper I for the stellar system.}
\label{tab:star_inf}
\begin{tabular}{cccc}
\hline
 & Star A & \multicolumn{2}{c}{StarB} \\ \hline
Models &  & M1 & M2 \\ \hline
Age (Gyr) & 10.21$\pm$2.71 & 15.77$\pm$4.11 & 10.68$\pm$0.481\\
$\rm [M/H]_0$ (dex) & -0.4599$\pm$0.1313 & 0.1087$\pm$0.0999 & -0.3712$\pm$0.0484\\
$M~(\rm M_\odot)$ & 0.8769$\pm$0.0951 & 0.8907$\pm$0.0800 & 0.8281$\pm$0.0146\\
$R~(\rm R_\odot)$ & 1.9732$\pm$0.1037 & 1.1601$\pm$0.0495 & 1.0920$\pm$0.0146\\
$T_{\rm eff}$ (K) & 5538.92$\pm$118.7 & 5465.82$\pm$11918 & 6044$\pm$41\\
$\log(g)$ (dex) & 3.7870$\pm$0.0052 & 4.2565$\pm$0.0048 & 4.2781$\pm$0.0016\\
$\rm [Fe/H]$ (dex) & -0.5140$\pm$0.1382 & 0.0467$\pm$0.1038 & -0.4503$\pm$0.0507\\
$L~(\rm L_\odot)$& 3.3343$\pm$0.5055 & 1.090$\pm$0.1557 & 1.2805$\pm$0.0252 \\
$\rm \Delta\nu\ (\mu Hz)$ & 45.62$\pm$1.17 & 102.00$\pm$2.05 & 107.74$\pm$0.62\\
$\rm \nu_{max}\ (\mu Hz)$ & 708.53$\pm$3.70 & 2097.94$\pm$3.05 & 2097.80$\pm$3.06\\
$Z_i$ & 0.0047$\pm$0.0015 & 0.0172$\pm$0.0039 & 0.0067$\pm$0.0006 \\
$Y_i$ & 0.2875$\pm$0.0279 & 0.2753$\pm$0.0254 & 0.2856$\pm$0.0051\\ \hline

\end{tabular}
\end{table*}

\section{Stellar models}
\label{sec:Stellar_models}

\subsection{Physical inputs}
\label{sec:phys_inp}

The stellar models of HD~81809 adopted in this work were computed using the \textsc{MESA} code \citep[][]{Paxton2011,Paxton2013,Paxton2015,Paxton2018,Paxton2019}, adopting the same input physics as in Paper~I for consistency. These correspond to the Grid~C physics presented in \cite{Moedas2025}, and are summarized below.  

We used the solar chemical mixture from \cite{Asplund2009} together with the OPAL opacity tables \citep{Iglesias1996} for high temperatures, and the low-temperature tables from \cite{Ferguson2005}. Atomic diffusion was included, neglecting radiative accelerations, following the \cite{Thoul} prescription implemented in \textsc{MESA}. The equation of state was taken from OPAL2005 \citep{Rogers2002}, and the atmospheric boundary condition was based on \cite{Krishna1966}. Convection was treated according to the mixing-length theory \citep{Cox1968}, with the mixing-length parameter ($\alpha_\mathrm{MLT}$) calibrated on the Sun.

The models also include turbulent mixing below the convective envelope, following the prescription of \cite{Proffitt1991}:
\begin{equation}
    D_\mathrm{T} = C \left( \frac{\rho_\mathrm{BC}}{\rho} \right)^n,
    \label{eq:Dturb}
\end{equation}
where $\rho$ is the local density, $\rho_\mathrm{BC}$ is the density at the bottom of the convective zone, and $C$ and $n$ are constants. We adopted $n = 1.3$ from \cite{Eggenberger2022} (should reproduce the rotation mixing of the sun), and $C=1615$ calibrated on solar models to reproduce the solar surface lithium abundance \citep{Moedas2025}.

\subsection{Accretion in stellar models}
\label{sec:accretion}

\begin{table}[]
\centering
\caption{Accretion rate and total accreted mass in models assuming accretion of pure metals.}
\label{tab:acre_models}
\begin{tabular}{ccccc}
\hline
Model & \begin{tabular}[c]{@{}c@{}}Accretion\\ Type\end{tabular} & $\Delta M~(\mathrm{M_\odot/yr})$ & $\Delta M_\mathrm{TAMS}~(\mathrm{\Meart})$ \\ \hline
M2 & None & 0 & 0 \\ \hline
CA1 & \multirow{3}{*}{Constant} & $10^{-14}$ & 38   \\
CA2 &  & $2\times10^{-14}$ & 76   \\
CA3 &  & $6\times10^{-14}$ & 228  \\ \hline
PAe1 & \multirow{3}{*}{Punctual} & \multirow{3}{*}{$10^{-11}$} & 150   \\
PAe2 &  &  & 200  \\
PAe3 &  &  & 250  \\ \hline
PAl1 & \multirow{3}{*}{Punctual} & \multirow{3}{*}{$10^{-11}$} & 25   \\
PAl2 &  &  & 50  \\
PAl3 &  &  & 75  \\ \hline
\end{tabular}
\end{table}

We simulated the engulfment of planetary material using the \texttt{mass\_change} input option in MESA\footnote{\url{https://github.com/nmoedas/Accretion-Routine.git}},
 which allows the stellar mass to vary over time. Positive values correspond to mass accretion, while negative values represent mass loss, with the mass change defined in units of solar masses per year. Accretion has been implemented in \textsc{MESA} since \cite{Paxton2011} and subsequent updates in \cite{Paxton2015,Paxton2019} improving energy conservation, guided by the analytic framework developed for white dwarfs by \cite{Townsley2004}. 
 Although MESA offers several options for specifying the chemical composition of the accreted material, we performed a series of controlled tests designed to isolate the impact of accretion rate, timing, and chemical mixture.

\begin{figure}[]
    \centering
    \includegraphics[width=1\linewidth]{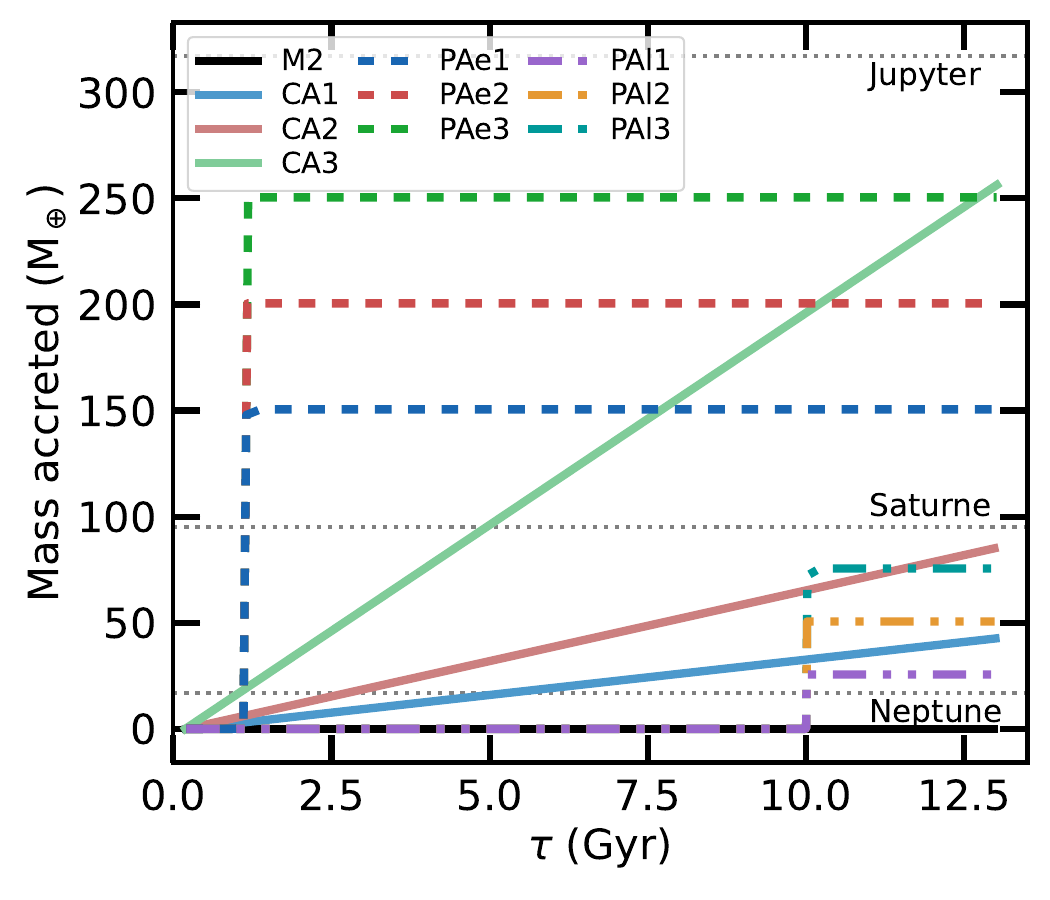}
    \caption{Mass accreted during the stellar evolution, horizontal gray dotted lines mark the masses of Jupyter, Saturn and Neptune.}
    \label{fig:mass_acre}
\end{figure}

\subsubsection{Accretion of pure metals}
As a first step, we adopted a simplified assumption in which the accreted material consists entirely of heavy elements (pure metals).
The metal ratios of the accreted material are
 assumed to be identical to those of the initial stellar composition, adopting
 a solar chemical mixture from  \cite{Asplund2009}.
Though this assumption may be unrealistic for  the engulfment of a Neptune-sized or larger planet, it allows to draw a base line
for estimating the amount of material required to reproduce the observed chemical variations, without introducing steep composition gradients.

First, we consider a scenario in which the star accretes the material slowly and steadily throughout its evolution, starting from the zero-age main sequence (ZAMS) (these models are labeled with CA).
Constant accretion produces a smooth chemical evolution, as diffusion and turbulent mixing operate simultaneously with the accretion process. To counteract atomic diffusion and reproduce the observed [Fe/H] of HD~81809~B, the star must accrete more than a Neptune-mass equivalent of metals over its lifetime, and up to Saturn or Jupiter-like metal 
masses depending on the adopted accretion rate (Table \ref{tab:acre_models}).

Secondly, we consider the engulfment to occur instantaneously (PA models). Since MESA does not allow strictly instantaneous mass injection,
we adopt a high accretion rate  of $10^{-11}~\rm (M_{\odot}/yr)$ 
with a maximum time step of $10^4~\rm yr$ between stellar models. This slows down the computation, but ensures  convergence stability.
Under these assumptions, accreting a Jupiter-mass planet ($\sim300~\Meart$)  occurs over $\simeq 100$ Myr, which is effectively instantaneous compared to stellar evolutionary timescales
(we will refer to it as a punctual event). Instantaneous engulfment can occur at different stages of stellar evolution. In this work, we consider accretion occurring either near the ZAMS (PAe models) or near the current stellar age (PAl models), when the star reaches 10~Gyr.
In this manuscript, we present three runs (for the PAe and PAl models) with different accreted masses that allow us to reproduce the $\FeH$ at the expected stellar age. 

The models are summarized in Table~\ref{tab:acre_models}, including the type of accretion, the adopted mass accretion rate, and the approximate mass variation at the terminal-age main sequence (TAMS). 
To reproduce the observed [Fe/H],  early accretion requires $\Delta M\geq150~\Meart$ of metals, whereas late accretion requires only $\Delta M\simeq(25-75)~\Meart$. 

Figure~\ref{fig:mass_acre} provides a quantitative comparison of the accreted masses with those of Solar System planets, showing the time evolution of accretion in all models.
The punctual accretion cases (PAe and PAl)  create sharp variations in stellar mass consistent with instantaneous engulfment events, while the constant accretion models exhibit a gradual mass increase.
For the constant accretion models, the total accreted mass ranges between Neptune-like and Saturn-like masses for the two lowest accretion rates (CA1 and CA2), and between Saturn-like and Jupiter-like masses for the highest rate. In the punctual accretion models, early-age events (PAe) correspond to accreted masses between those of Saturn and Jupiter, whereas late-age events (PAl), occurring near the expected stellar age, correspond to masses between those of Neptune and Saturn.

\begin{table}[]
\centering
\caption{Total mass accreted, fraction of accreted mass metals and gas envelope in models with hydrogen and helium accretion.}
\label{tab:acre_models_XY}
\resizebox{\linewidth}{!}{
\begin{tabular}{ccccc}
\hline
Model & $\Delta M_\mathrm{TAMS}~(\mathrm{\Meart})$ & $M(Z)~(\mathrm{\Meart})$ & $M(X+Y)~(\mathrm{\Meart})$ & $M(X+Y)\%$ \\ \hline
PR1 & \multirow{6}{*}{75} & 75 & 0 & 0 \\
PR2 &  & 56.25 & 18.75 & 25 \\
PR3 &  & 37.5 & 37.5 & 50 \\
PR4 &  & 18.75 & 56.25 & 75 \\
PR5 &  & 3.75 & 71.25 & 95 \\
PR6 &  & 0.75 & 74.25 & 99 \\ \hline
PV1 & 50 & \multirow{3}{*}{50} & 0 & 0 \\
PV2 & 75 &  & 25 & 33 \\
PV3 & 300 &  & 250 & 75 \\ \hline
\end{tabular}
}
\end{table}

\subsubsection{Accretion including hydrogen and helium}

Since Neptune-like and larger planets contain substantial H–He envelopes, we next investigated the effect of accreting material with different metal-to-gas ratios.
Hereafter, we refer to the accreted material of hydrogen and helium elements as the gas envelope. Here we simulate two accretion events occurring close to the expected age. 
  
In the first event, we fixed the total mass accreted at $\Delta M_{\rm TAMS}=75~\Meart$ 
and varied the metal fraction $M(Z)$ from 100\% metals down to 1\% (PR models).

For the second event, we simulated the accretion of the same metal mass of $M(Z)=50~\Meart$, but with different gas envelopes (PV models). This  test is intended to mimic the engulfment of planets with atmospheres of different sizes. It is important to note that in both events, the hydrogen-to-helium ratio in the accreted gas is assumed to be the same as the initial stellar model.

The models for both tests are presented in Table~\ref{tab:acre_models_XY}. The adopted accretion rates are the same as those used in the pure-metal models.

\subsubsection{Thermohaline convection}

\begin{table}[]
\centering
\caption{Accretion rate and total accreted mass in models with thermohaline convection.}
\label{tab:acre_models_TH}
\begin{tabular}{ccccc}
\hline
Model & \begin{tabular}[c]{@{}c@{}}Accretion\\ Type\end{tabular}& $\Delta M~(\mathrm{M_\odot/yr})$  & $\Delta M_\mathrm{TAMS}~(\mathrm{\Meart})$ \\ \hline
TH1 & \multirow{4}{*}{Punctual} & \multirow{4}{*}{$10^{-11}$} & 50 \\
TH2 &&& 6 \\
TH3 &&& 250 \\
TH4 &&& 50 \\  \hline
TH5 &Constant&$6\times10^{-14}$& 226 \\  \hline
\end{tabular}
\end{table}

\cite{Soares2025} point out that chemical variations  caused by planet engulfment may occur in stars, but their observational signatures are rare since chemical transport mechanisms, such as atomic diffusion and turbulent mixing, can erase these variations. Both of these processes are included in the input physics of our stellar models, as described in Sec.~\ref{sec:phys_inp}.
However, all of the models discussed so far neglect the thermohaline convection, a chemical mixing mechanism that affects stellar evolution on a shorter timescale. 
Large chemical gradients, which could be produced by planet engulfment, trigger thermohaline instabilities that rapidly homogenize the affected region over a short period of time. \cite{Deal2015} showed that in 16 Cyg B, small accretion of less than one Earth mass is sufficient to create thermohaline convection, which depletes lithium without altering the composition of other elements. For the tests performed in this manuscript, thermohaline convection is not negligible. To asses its impact, we performed an additional set of models with thermohaline convection activated (TH models), adopting the prescription of  \cite{Brown2013}, which is already implemented in MESA.

 The models that consider thermohaline convection are presented in Table \ref{tab:acre_models_TH}. In models TH1 and TH2 the accretion event occurs near the star's current age, in TH3 and TH4 it occurs at an early stage of stellar evolution. Model TH5 considers a constant accretion rate, similar to model CA3. In all TH models, the accreted material is  assumed to consist only of metals.

\subsubsection{Accretion with a rocky (CI-Chondrite) mixture}

The previous models assumed solar-scaled metal ratios. However, the chemical composition of planets does not necessarily reflect the same elemental ratios as their host stars \citep{Adibekyan2018}. In addition, planetary composition depends on the formation location within the protoplanetary disk, as discussed by \cite{Soares2025}. Planets formed closer to the host star are typically enriched in refractory elements such as magnesium, silicon, and iron, whereas planets formed farther away tend to contain higher fractions of volatile elements.

To consider a more realistic rocky engulfment scenario for HD~81809, we can adopt a CI-chondrite–based composition.  We use the meteoritic abundances presented in \cite{Asplund2009}, based on the results by \cite{Lodders2009}. 
These abundances are provided in the customary astronomical logarithmic scale: 
\begin{equation}
\log(\epsilon_\texttt{A})=\log\left(\frac{N_\texttt{A}}{N_H}\right)+\log(\epsilon_H),
\end{equation}
where $N_\texttt{A}$, $N_H$ is the number densities of elements $\texttt{A}$ and hydrogen and by definition $\log(\epsilon_H)=12$.

To implement this  accretion mixture in MESA we can use the \texttt{z\_fraction\_A} option,  which allows specifying the mass fraction of each individual \texttt{A} element. We converted the 
the logarithmic abundances into mass fractions 
using:
\begin{equation}
    \log\left(\frac{Z(\texttt{A})}{X}\right)=\log\left(\frac{N_\texttt{A}}{N_H}\right)+\log\left(A_t(\texttt{A})\right),
\end{equation}
in which $Z(\texttt{A})$ and $A_t(\texttt{A})$ are the mass fraction and atomic weight of element $\texttt{A}$, and $X$ is the hydrogen mass fraction.
The resulting mass fractions adopted are: Li=1.5$\times10^{-6}$ C=0.0341856, N=0.0029553, O=0.4659682, Mg=0.0954898, Al=0.0084205, Si=0.0525079, S=0.1053748, Ca=0.0090609, and Fe=0.1824933.  It is important to note that these
values 
correspond to Solar System meteoritic material and may not accurately represent the composition of rocky bodies in the HD~81809 system.
 
 We performed two simulations using this meteoritic mixture, assuming  accretion of 6 and 50 $~\Meart$. 
  Additionally, to investigate the impact of lithium enrichment, we carried out two additional simulations for the $50~\Meart$ case, reducing the lithium mass fraction by factors of $10$ and $100$. These models are summarized in Table~\ref{tab:acre_models_diff_mixture}.

\begin{table}[]
\centering
\caption{Mass fraction of lithium, accretion rate and total accreted mass of models with CI-Chondrite mixture.}
\label{tab:acre_models_diff_mixture}
\begin{tabular}{ccccc}
\hline
Model & Li& $\Delta M~(\mathrm{M_\odot/yr})$  & $\Delta M_\mathrm{TAMS}~(\mathrm{\Meart})$ \\ \hline
DM1 & \multirow{2}{*}{$1.5\times10^{-6}$} & \multirow{2}{*}{$10^{-11}$} & 6 \\
DM2 &&& 50\\\hline
DM3 & {$1.5\times10^{-7}$} & {$10^{-11}$} & 50 \\\hline
DM4 & {$1.5\times10^{-8}$} & {$10^{-11}$} & 50 \\\hline
\end{tabular}
\end{table}

\begin{figure}[]
    \centering
    \includegraphics[width=1\linewidth]{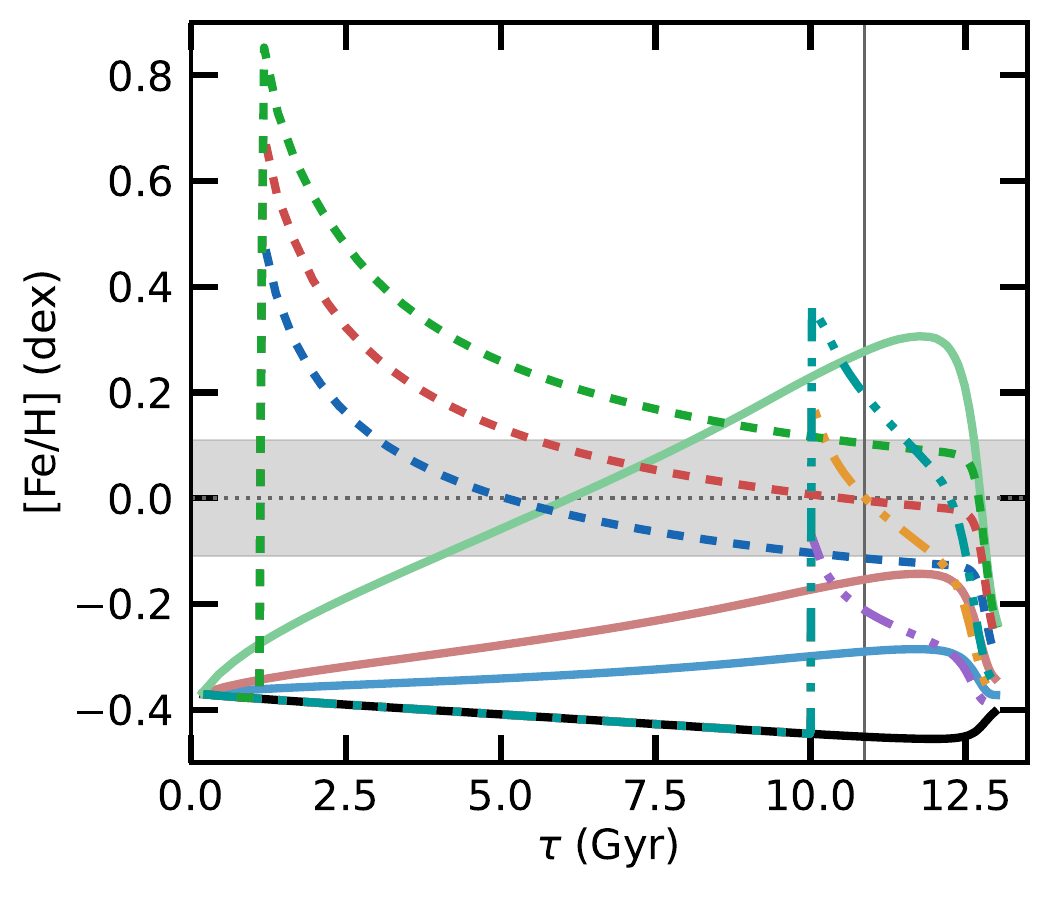}
    \includegraphics[width=1\linewidth]{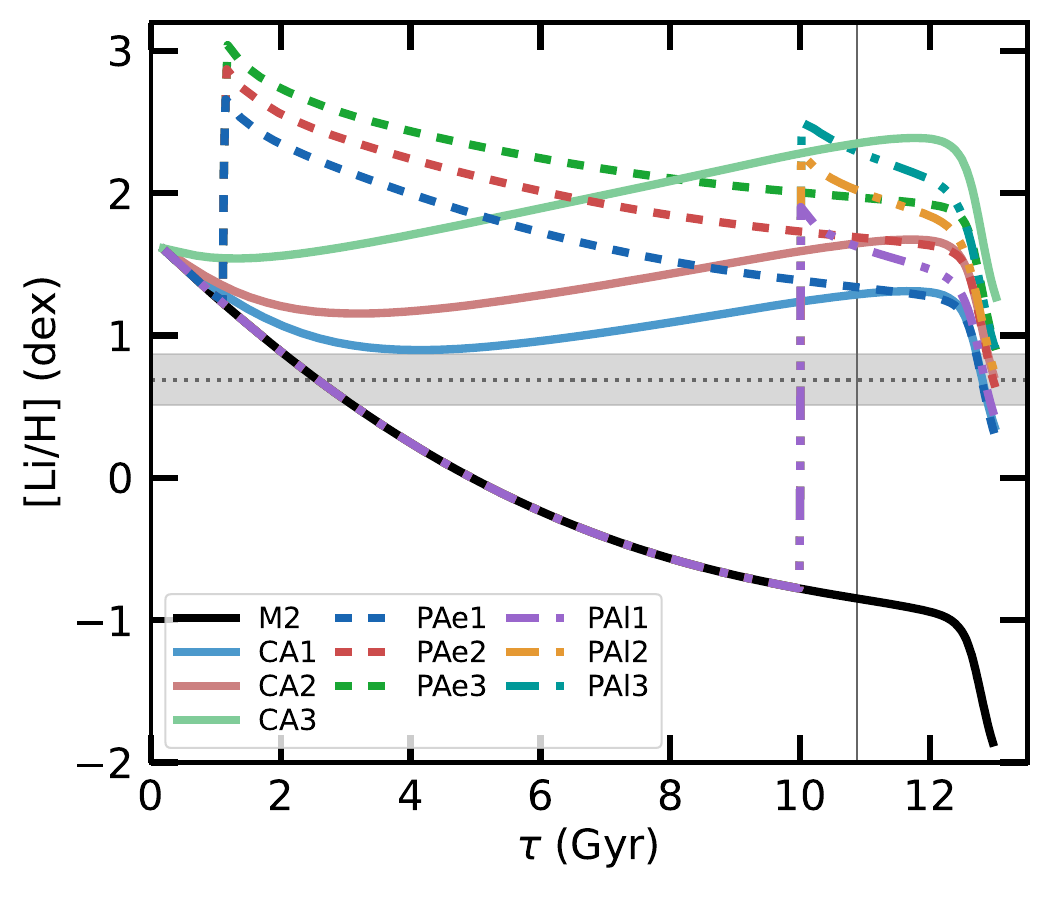}
    \caption{Top panel surface evolution of [Fe/H], horizontal dotted gray line is the observed [Fe/H], the vertical line is the expected age for the inferred best model. bottom  panel surface evolution of [Li/H], horizontal dotted gray line is the observed [Li/H], the vertical line is the expected age for the inferred best model. Models described in tables~\ref{tab:acre_models}.}
    \label{fig:acre_pure_met}
\end{figure}

\begin{figure}[]
    \centering
    \includegraphics[width=1\linewidth]{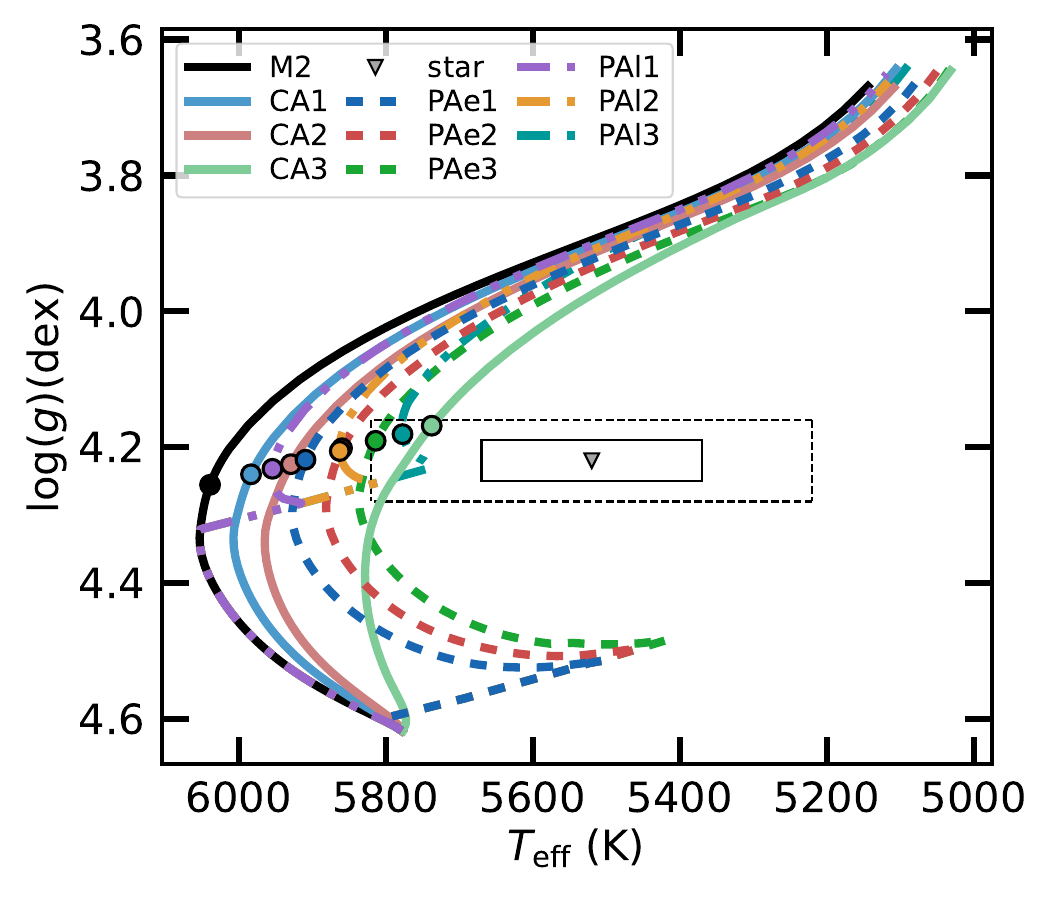}
    \includegraphics[width=1\linewidth]{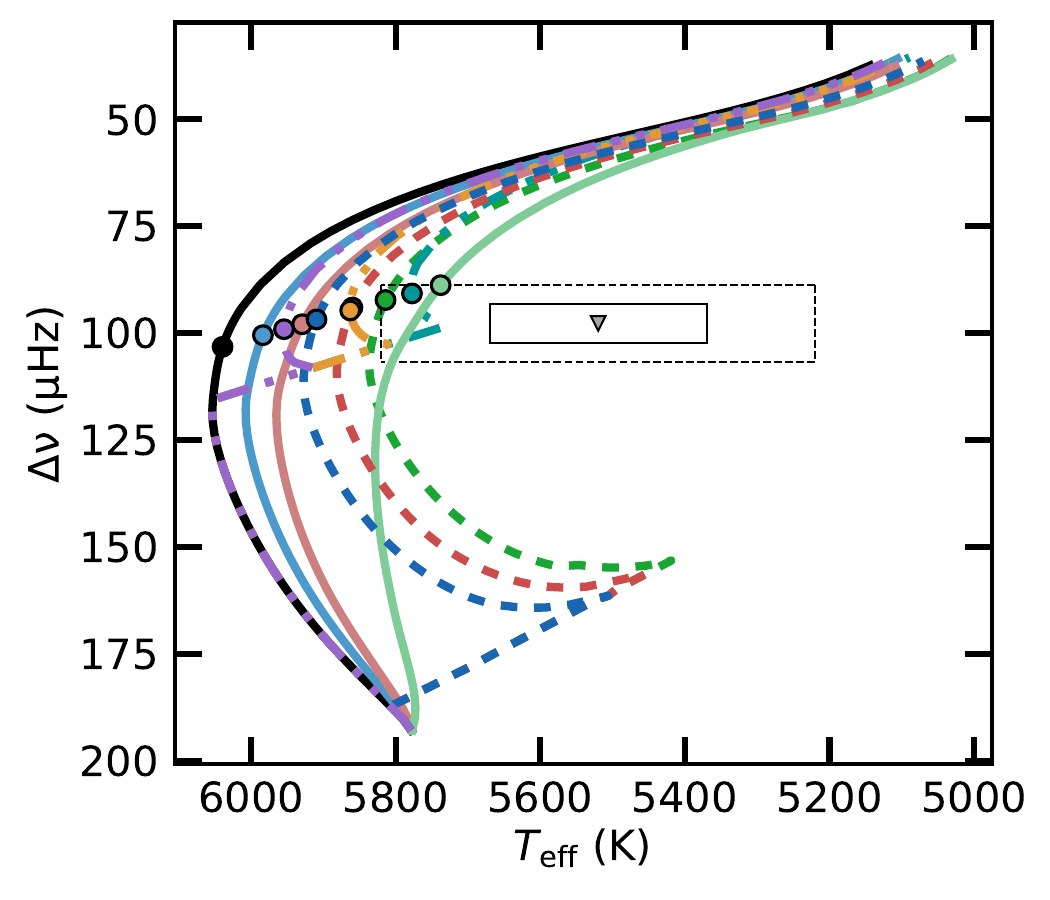}
    \caption{Kiel diagram (top panel) and asteroseismic diagram for $\Delta\nu$ (bottom panel), showing the evolutionary tracks of the different models considering accretion. The triangle indicates the observed values with relative 1$\sigma$ and 2$\sigma$ uncertainty boxes. The circles indicate the model with the current age of HD~81809B for each evolutionary track.}
    \label{fig:HR_after_acre}
\end{figure}

\section{Results of stellar accretion simulations}
\label{sec:accre_models}
\subsection{Accretion of pure metal material}

In our first set of models, we consider that the star accretes pure metal material through either constant accretion or punctual engulfment events.
We specifically test a range of accreted metal-rich (rocky) material, taking into account previous planetary core estimates. \cite{Sato2005} modeled the giant planet HD~149026b and inferred a core mass of $\sim 67~\Meart$ in heavy elements, as confirmed by \cite{Wolf2007}. More recently,
 \cite{Yildiz2024}
  used MESA to model Jupiter, Saturn, Neptune, and Uranus, estimating core masses of approximately 40, 25, 14, and 12 $\Meart$, respectively.
  This suggests that accretion masses  explored in our late-age models are physically plausible, representing the engulfment of the rocky core of a giant planet whose gaseous envelope had been partially or fully removed through evaporation or tidal stripping.

Figure~\ref{fig:acre_pure_met} shows the evolution of surface iron and lithium abundances during accretion. 
Constant accretion produces a relatively smooth chemical evolution with no abrupt variations. For the lower accretion rate (CA1), the stellar $\FeH$ remains close to its ZAMS value, indicating that accreting more than a Neptune-like mass is required to overcome atomic diffusion effects. To reproduce the $\FeH$ enrichment observed in HD~81809B, the star must accrete a mass between that of Saturn and Jupiter, assuming constant accretion with atomic diffusion and turbulent mixing active.

Surface lithium evolution exhibits a different behavior. 
We notice that across all simulated cases, the final surface lithium abundance, [Li/H], remains systematically higher than observed value. Moreover, all model undergo an initial phase of lithium depletion during the early Main Sequence (MS), before exhibiting a surface enrichment. This suggests that
while the depletion mechanisms are highly efficient in early evolutionary stages, 
they lose efficacy as the star evolves.

For punctual accretion events, surface abundances show a sharp variation when the engulfment occurs. The required accreted mass to reproduce the observed $\FeH$ strongly depends on
the timing of the event: early accretion requires $\sim 150~\Meart$ of metals to reach the observed $\FeH$ by the estimated age, whereas later accretion events need only a mass variation between $\sim 25$–$75~\Meart$.
Despite the timing, also all the punctual events lead to excessive lithium enrichment at the surface.

The Li overabundance shown by all the models with accretion of pure metal material may be the result of having neglected thermohaline convection, which can mix the accreted material and reduce surface lithium enrichment.

It is important to assess how accretion events affect the star's surface parameters.
 From Paper I, we know that
 models with low initial metallicity
 failed to
  reproduce the surface temperature. Figure \ref{fig:HR_after_acre}  shows the Kiel and seismic diagrams for the evolutionary tracks of the different accretion tests. 
  Accretion increases the surface iron abundance, which enhances the stellar opacity and consequently reduces the effective temperature, $\Teff$. Models with larger accreted masses can approach within 2$\sigma$  the observed $\Teff$. 
   We also notice that 
   models matching the estimated stellar age
   show an increase in $\log(g)$ and $\Delta\nu$, moving away from the observed values, though still within 2$\sigma$. This behavior
   could explain why, 
   standard models of metal-poor stars tend to predict higher effective temperatures. Including accretion events that raise $\FeH$ can therefore be necessary to reproduce the observed $\Teff$, suggesting that such events should be incorporated in stellar models in order to reproduce the observed parameters.

\begin{figure*}[]
    \centering
    \includegraphics[width=\linewidth]{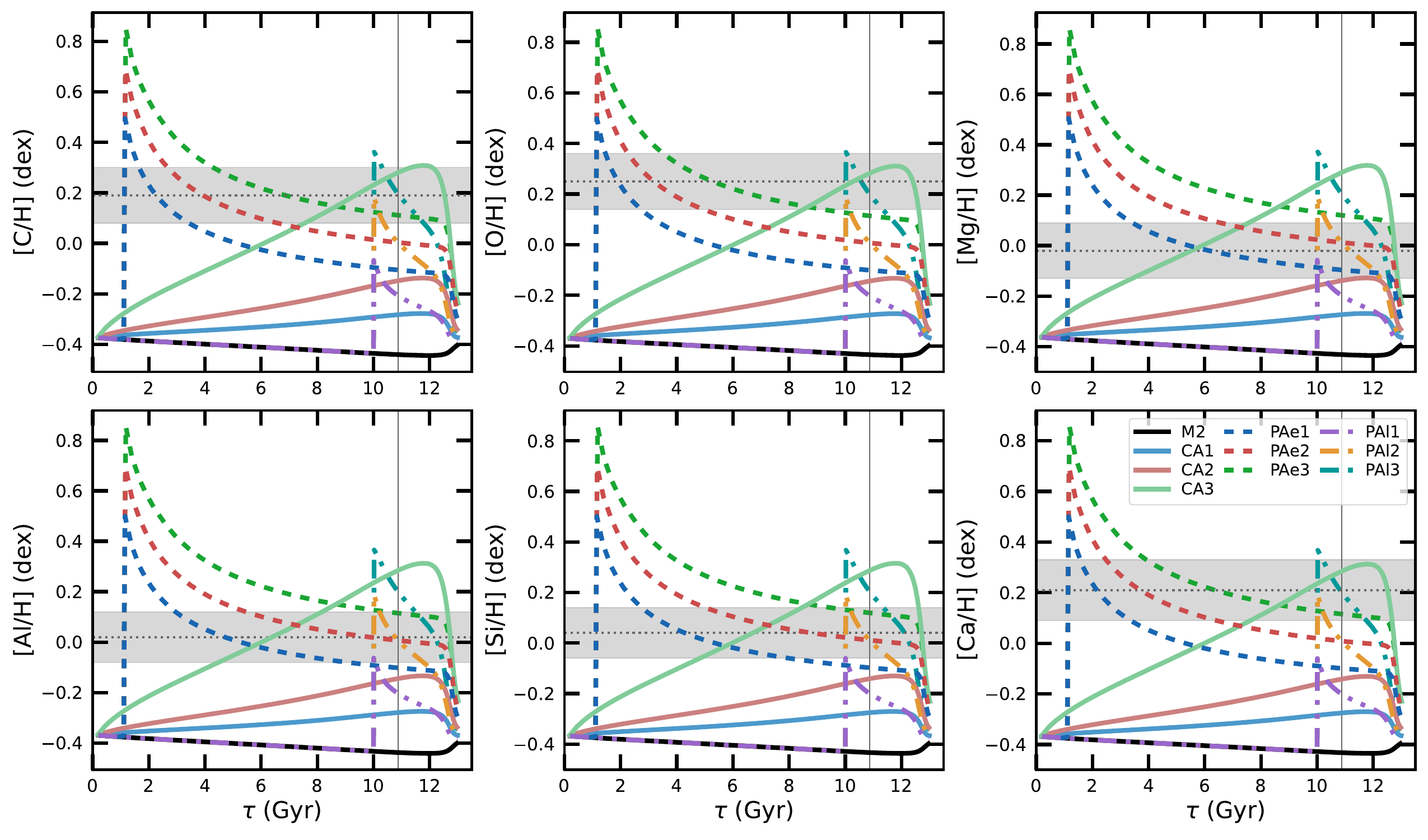}
    \caption{The top panels show the evolution of the surface abundances [C/H], [O/H], and [Mg/H], while the bottom panels show [Al/H], [Si/H], and [Ca/H] for the first set of accretion models. The horizontal dotted gray line shows the observed abundance each element from Paper~I, and the vertical line shows the age of the best-fit stellar model.}
    \label{fig:accret_other_elements}
\end{figure*}

It is also important to examine how accretion affects the surface abundances of other elements.
Figure \ref{fig:accret_other_elements} shows the evolution of carbon, oxygen, magnesium, aluminum, silicium, and calcium. 
These elements exhibit trends broadly similar to that of iron, with surface abundances increasing following accretion. However, different elements require different accreted masses to reproduce their observed abundances. Magnesium, aluminum, and silicon can be matched with accretion masses comparable to those inferred from iron, whereas carbon, oxygen, and calcium require larger accreted masses.

This discrepancy does not rule out an accretion scenario, but instead suggests that the composition of the accreted material and/or the initial stellar composition differs from the assumed mixture. In particular, it indicates that the engulfed material may not follow a solar-scaled abundance pattern, but instead reflect a chemically differentiated rocky body.

\begin{figure}[]
    \centering
    \includegraphics[width=1\linewidth]{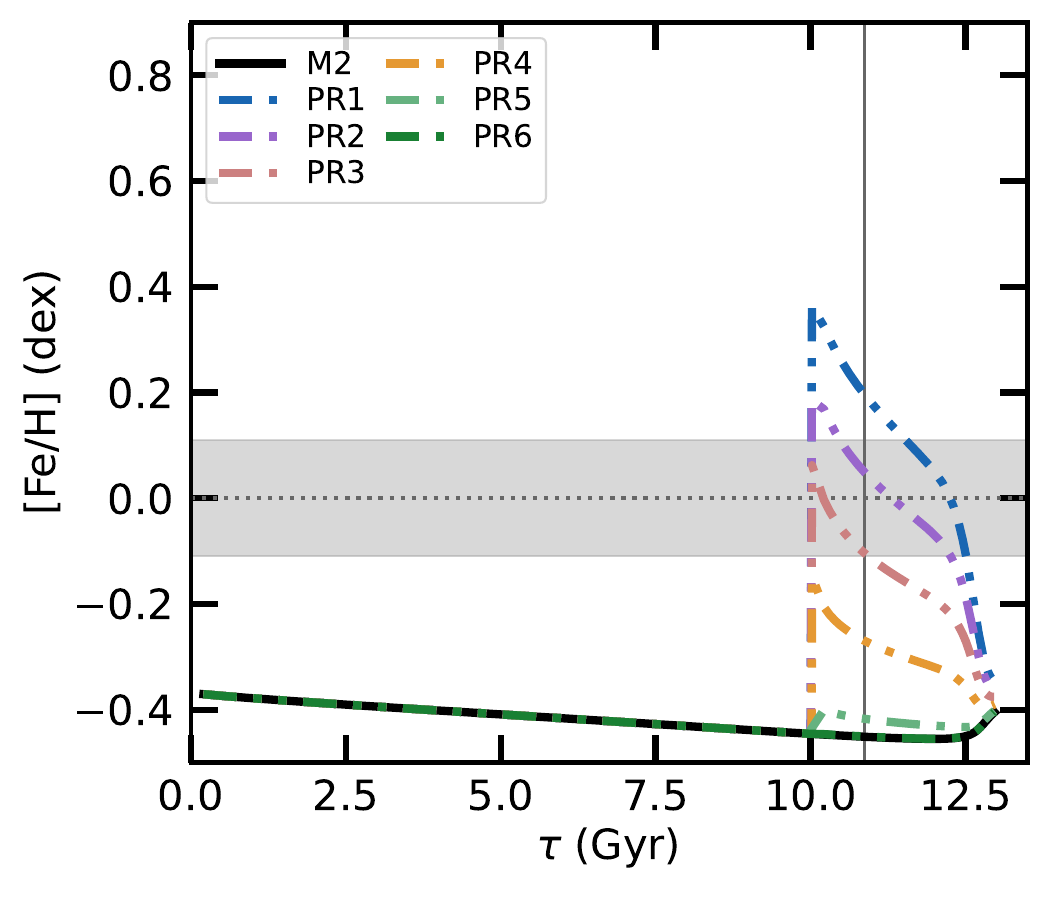}
    \includegraphics[width=1\linewidth]{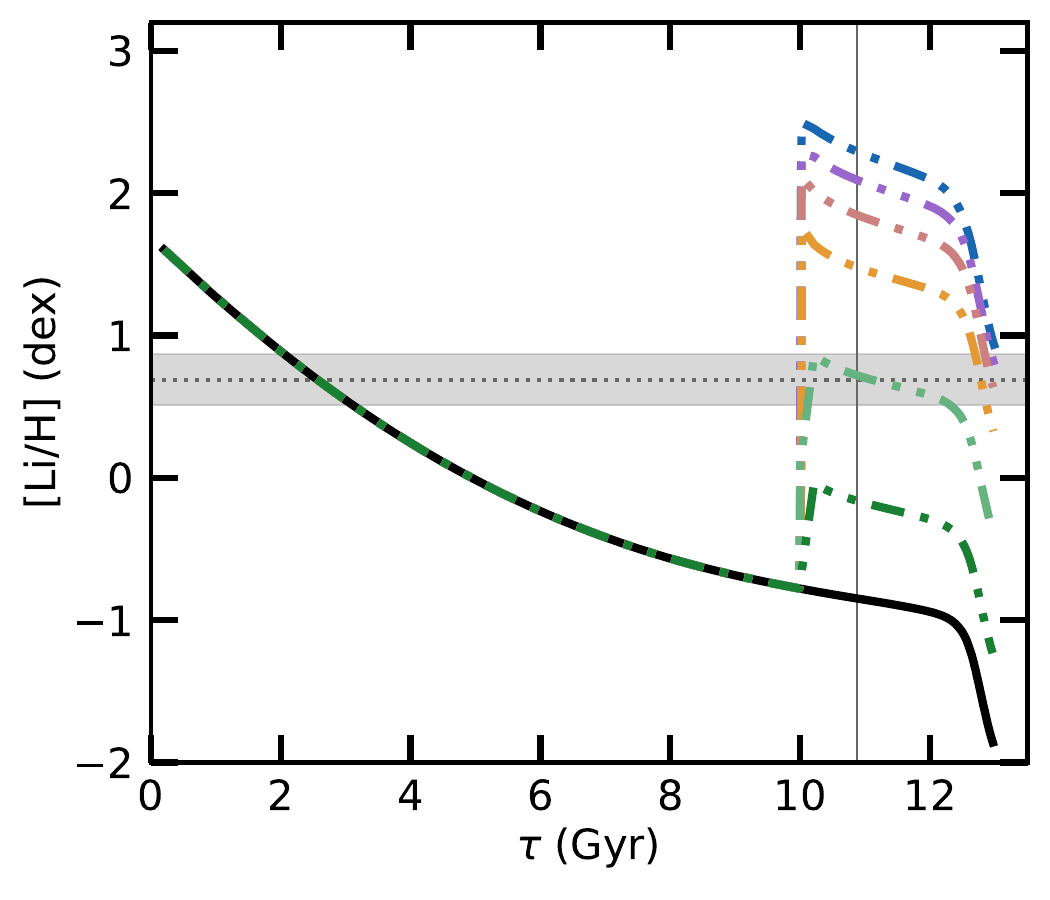}
    \caption{Same plots as Fig. \ref{fig:acre_pure_met}, but considering accretion event of 75 $\Meart$ planet with different ratios between metal and hydrogen and helium composition. Models described in tables~\ref{tab:acre_models_XY}.}
    \label{fig:acret_diff_ratios_XY}
\end{figure}

\subsection{Accretion with hydrogen and helium}

The engulfment of a Neptune-mass or larger planets is expected to involve not only metals
but also significant amounts of hydrogen and helium from the planetary envelope.
It is therefore important to assess how the metal-to-gas ratio of the accreted material affects the stellar surface composition. 

In this test, we consider different metal-to-gas ratios while keeping the total accreted mass fixed at $75~\Meart$.
Figure~\ref{fig:acret_diff_ratios_XY} shows the evolution of surface $\FeH$ and $\rm[Li/H]$ for these models. As expected, decreasing the metal fraction leads to smaller surface enrichment. In the most extreme case (PR6), where only 1\% of the accreted mass consists of metals, the evolution of $\FeH$ remains nearly indistinguishable from the non-accreting model. This means that  the accreted material produce a negligible change in the surface metallicity.

Lithium shows a similar qualitative trend, with lower metal fractions producing weaker surface enrichment. However, unlike iron, the PR6 model exhibits a noticeable increase in $\rm[Li/H]$ at the time of accretion. This occurs because the accreted material contains lithium at levels closer to the primordial stellar abundance, which is significantly higher than the lithium abundance in the evolved stellar surface. As a result, even a small amount of accreted lithium can produce a measurable increase in surface $\rm[Li/H]$.

These results demonstrate that lithium is particularly sensitive to accretion events, even when the accreted material is dominated by hydrogen and helium. This sensitivity makes lithium a powerful diagnostic of recent planetary engulfment.

To assess the impact of gaseous envelopes on the chemical signatures of accretion, we performed a second set of tests in which the star accretes a fixed metal mass of $50~\Meart$, combined with hydrogen–helium envelopes of varying mass, ranging from no envelope up to $250~\Meart$ (PV models). In these scenarios, increasing the envelope mass reduces the overall metal fraction of the accreted material, effectively diluting the heavy elements.

\begin{figure}[t]
    \centering
    \includegraphics[width=1\linewidth]{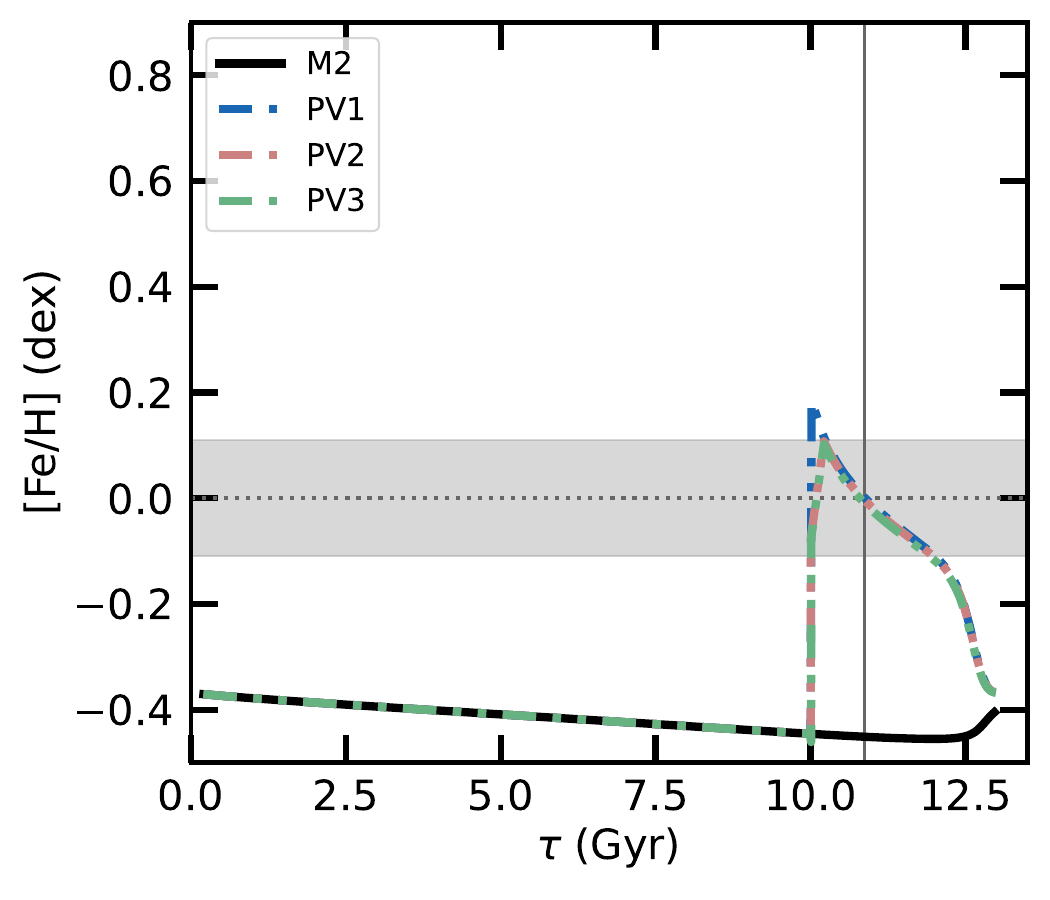}
    \includegraphics[width=1\linewidth]{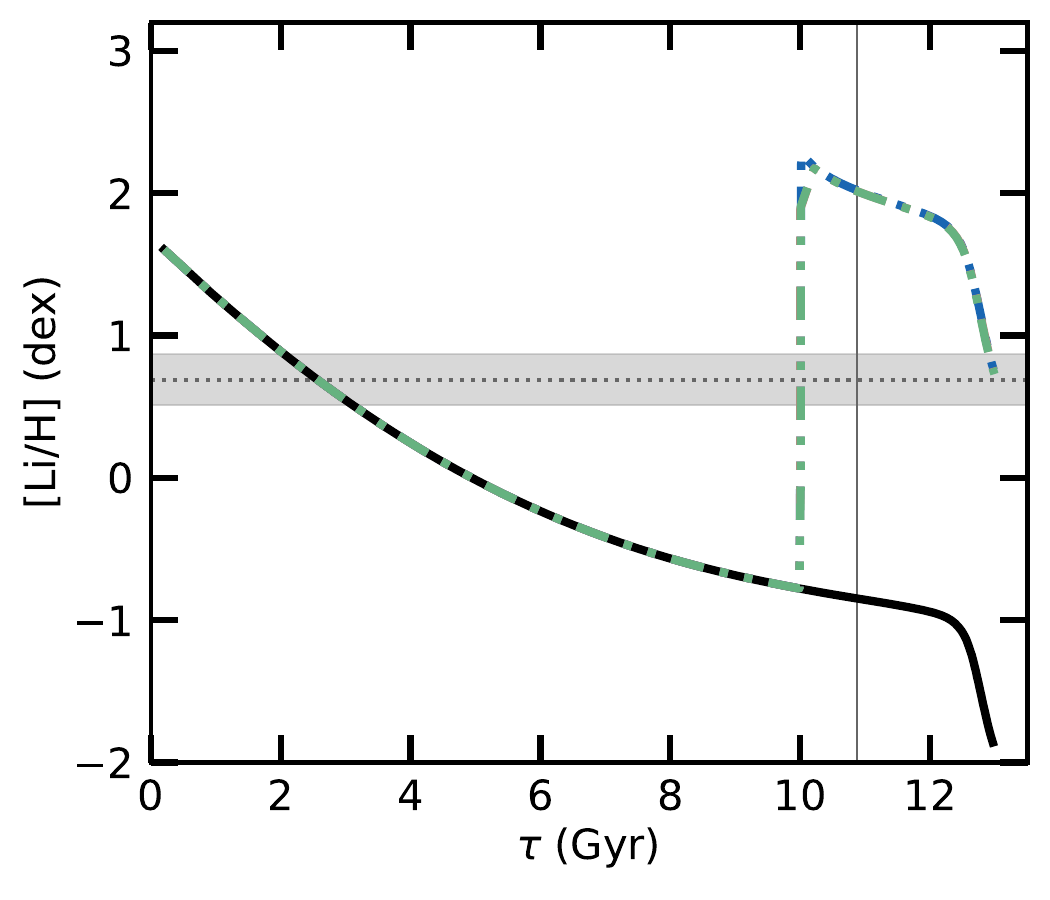}
    \caption{Same plots as Fig. \ref{fig:acre_pure_met}, but considering accretion event of metal core of 50 $\Meart$ with different mass  of hydrogen and helium envelops. Models described in tables~\ref{tab:acre_models_XY}.}
    \label{fig:acret_diff_env_XY}
\end{figure}

To assess the impact of gaseous envelopes on the chemical signatures of accretion, we performed a second set of tests in which the star accretes a fixed metal mass of $50~\Meart$, combined with hydrogen–helium envelopes of varying mass, ranging from no envelope up to $250~\Meart$ (PV models). In these scenarios, increasing the envelope mass reduces the overall metal fraction of the accreted material, effectively diluting the heavy elements.
Figure~\ref{fig:acret_diff_env_XY} shows the evolution of $\FeH$ and $\rm[Li/H]$ for these models. We find that increasing the envelope mass has a negligible effect on the final surface stellar abundance, which remain nearly identical to the case without a gaseous envelope. This indicates that the surface enrichment is primarily determined by the total mass of metals accreted, rather than by the mass of gas engulfed.

This could be because heavier elements are trace elements in the stellar interior and represent a very small fraction of the star's composition. As a result, the accretion of even modest amounts of metals produces a measurable change
in their surface abundances, whereas the addition of hydrogen and helium—already the dominant components of the stellar envelope—has little impact on the metal-to-hydrogen ratio.
 
 Only when the accreted metals are diluted within a sufficiently massive gaseous envelope, comparable to several thousand Earth masses (e.g., $\sim 4950~\Meart$ for $50~\Meart$ of metals) ,  
 it is expected to result in less significant changes to the stellar chemical composition. 
These results demonstrate that, for planet-size accretion events, the observable chemical signatures depend primarily on the total mass of metals accreted, rather than on the total mass of the engulfed planet. This reinforces the importance of rocky cores and metal-rich material as the dominant contributors to stellar chemical enrichment during planetary engulfment.

\begin{figure}[t]
    \centering
    \includegraphics[width=1\linewidth]{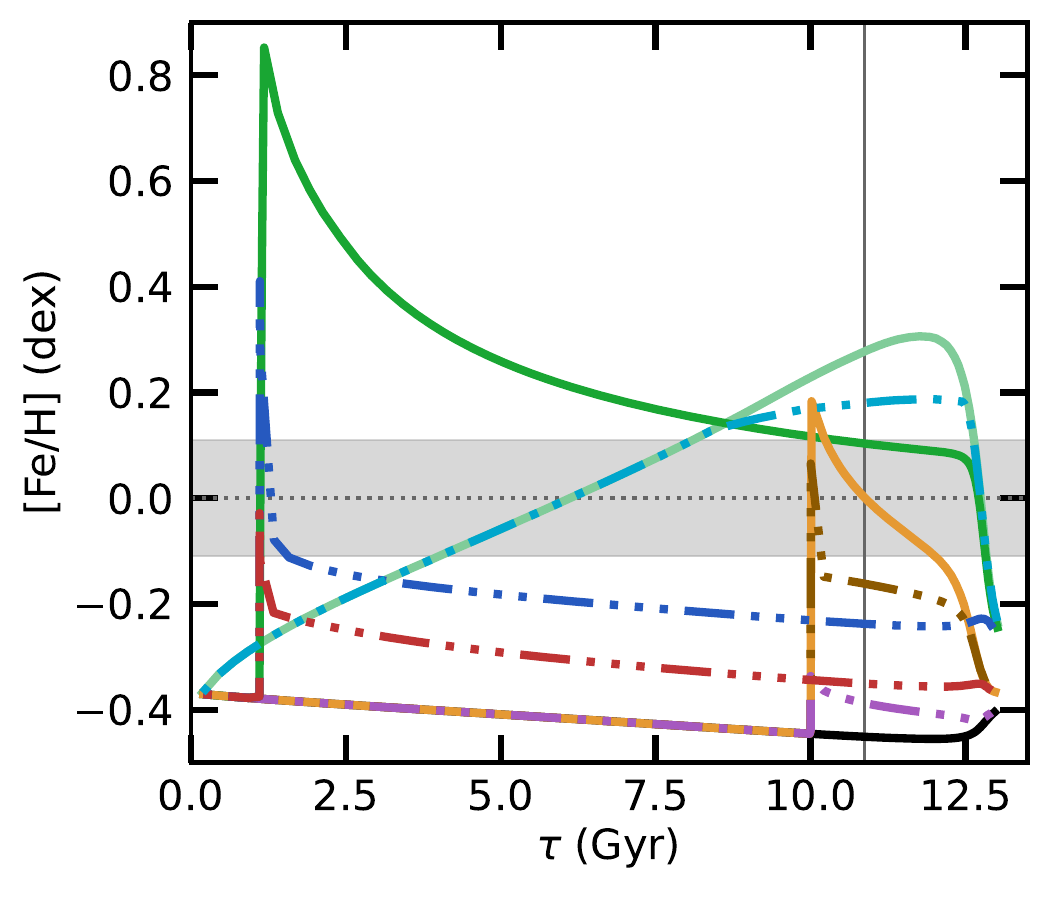}
    \includegraphics[width=1\linewidth]{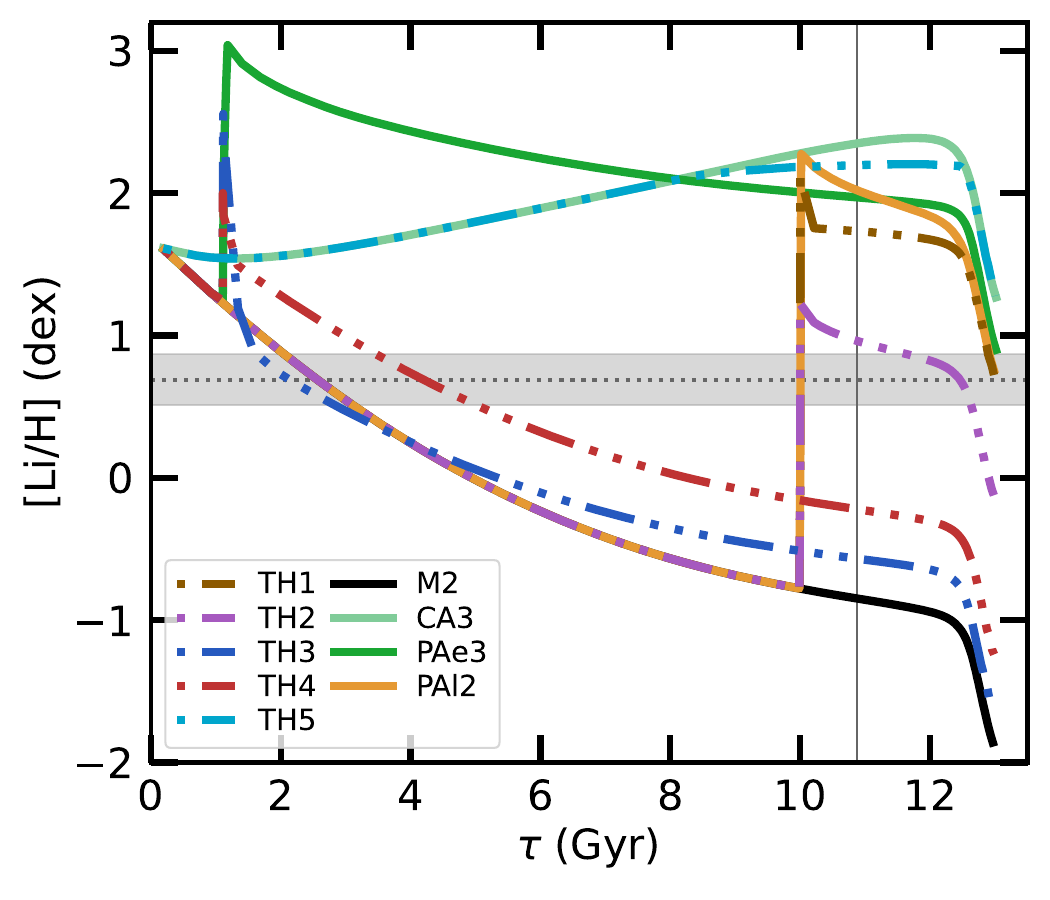}
    \caption{Same plots as Fig. \ref{fig:acre_pure_met}, comparison of accretion events with consideration of thermohaline convection. Models described in tables~\ref{tab:acre_models}~and~ \ref{tab:acre_models_TH}.}
    \label{fig:acret_thermohaline}
\end{figure}

\subsection{Impact of thermohaline convection}

To evaluate the effect of thermohaline convection on surface abundances after an accretion event, we focus on the accretion of heavy elements, since the inclusion of hydrogen and helium does not change the conclusions, as shown in the previous section. Figure~\ref{fig:acret_thermohaline} compares the surface evolution of $\FeH$ and $\rm[Li/H]$ for models with and without thermohaline convection. 

In the case of a $50~\Meart$ accretion event occurring near the present stellar age (PAL2 and TH1 models), thermohaline convection mitigates the chemical enrichment. 
The accreted material induces a steep mean molecular weight gradient that drives thermohaline mixing, leading to a rapid homogenization of the envelope and a significant decrease in $\mathrm{[Fe/H]}$ occurring within $\sim 200$~Myr. After this short phase, the depletion rate gradually approaches the value expected in models without thermohaline convection. Therefore, in order to match the observed $\mathrm{[Fe/H]}$ in the presence of thermohaline mixing, the accreted mass must be slightly higher than $50~\mathrm{M}_\oplus$.

For lithium, a similarly enhanced depletion is observed over the same timescale (about 200~Myr). However, this effect is still insufficient to reduce the lithium abundance to the values inferred from observations. We therefore explored additional accretion scenarios and found that, if the accretion event occurs close to the current stellar age, reproducing the observed lithium abundance requires the star to accrete less than 6~$\mathrm{M}_\oplus$ (model TH2), as shown in Fig.~\ref{fig:acret_thermohaline}.

We next consider models in which accretion occurs at the early stage of stellar evolution.
 For an accretion of 250~$\Meart$ (models TH3 and PAe3), the inclusion of
 thermohaline convection produces a dramatic effect on the surface chemical evolution.
 Thermohaline convection, triggered by the accreted material, quickly homogenizes the chemical composition over a timescale of about 200~Myr. 
 In the case of lithium (see lower panel of Fig. \ref{fig:acret_thermohaline}), the depletion is significantly enhanced, reaching values closer to those predicted by the model without  accretion. With a smaller mass accretion of 50~$\Meart$ (model TH4), the thermohaline convection has less effect, resulting in a higher surface lithium abundance.

 This demonstrates that the impact of thermohaline mixing depends on the  amount of accreted mass. 
 Larger accretion events produce steeper chemical gradients, which drive more vigorous thermohaline currents capable of penetrating deeper into the stellar interior. In the case of lithium, the enhanced depletion is therefore likely not only due to surface homogenization of this element, but also to the transport of material toward deeper layers where lithium is more efficiently destroyed.

For completeness, we also examine the impact of thermohaline convection in models with continuous accretion. We tested the high constant accretion rate ($\Delta M = 6\times10^{-14}~\Msun/\mathrm{yr}$, models CA3 and TH5) and found that the thermohaline convection has little effect on the stellar evolution. This occurs because gradual accretion produces only small chemical inhomogeneities. By the time a sufficiently steep chemical gradient  develops to trigger thermohaline mixing, atomic diffusion and turbulent mixing have had enough time to act, limiting the impact of thermohaline convection.

\begin{figure}[t]
    \centering
    \includegraphics[width=1\linewidth]{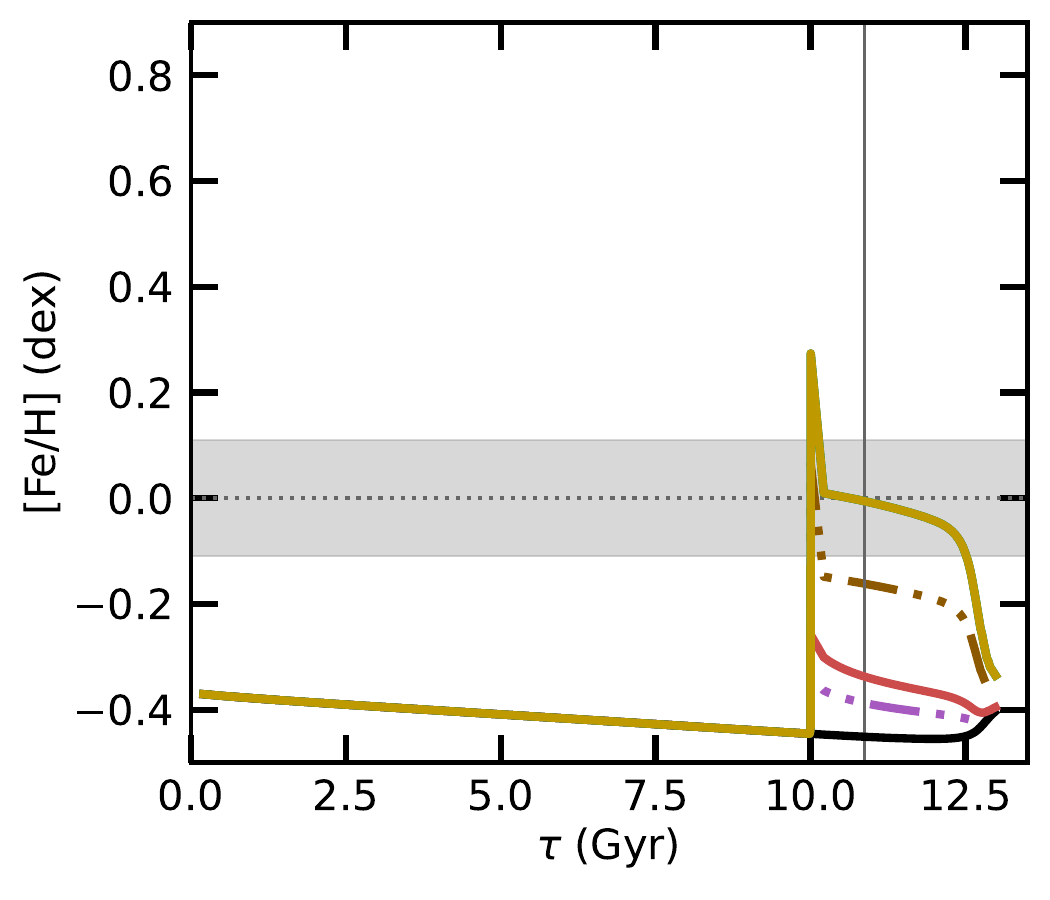}
    \includegraphics[width=1\linewidth]{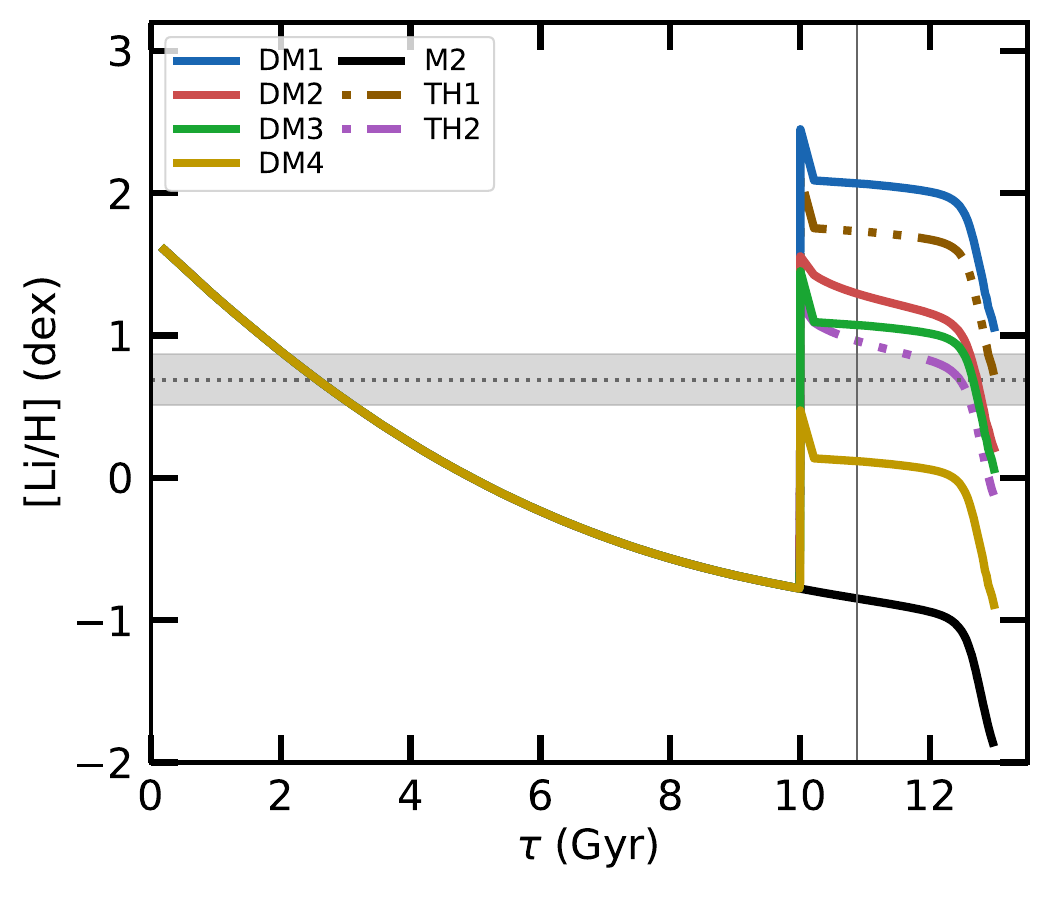}
    \caption{Same plots as Fig. \ref{fig:acre_pure_met}, comparison of accretion events with consideration of a different chemical mixture for the accreted material. Models described in tables \ref{tab:acre_models_TH} and \ref{tab:acre_models_diff_mixture}. }
    
    \label{fig:acret_diff_mixture}
\end{figure}

\subsection{Accretion of CI-Chondrite chemical mixture}

Previous tests revealed that in order to reproduce the observed $\FeH$ requires over-enriching the stellar surface in lithium. One possible explanation is that the assumed chemical composition of the accreted material is not representative. To test this, we considered a CI-chondrite mixture as described by \citep{Lodders2009}
(DM models) for meteorites. Figure \ref{fig:acret_diff_mixture} shows the resulting surface evolution of $\FeH$ and [Li/H] for an accretion event using this alternative chemical mixture.

First, we compare the $\FeH$ evolution of the DM models with the TH models. We observe that, for the same accreted mass, the stellar surface is more enriched in iron when using the new chemical mixture. This can be explained by the fact that the new CI-Chondrite mixture has a higher iron mass fraction. Therefore, 
a smaller amount of material is sufficient to reach the same surface enrichment.

The DM1 and DM2 models exhibit a larger variation in lithium compared to the TH1 and TH2 models, reflecting the higher lithium content in the meteorite mixture relative to the initial solar composition.
 To reproduce the observed lithium abundance for an accretion of 50~$\Meart$, the fraction of lithium in the accreted material needs to be reduced by a factor 10-100 (models DM3 and DM4) relative to  the CI-Chondrite composition (model DM1).

\begin{figure*}[]
    \centering
    \includegraphics[width=\linewidth]{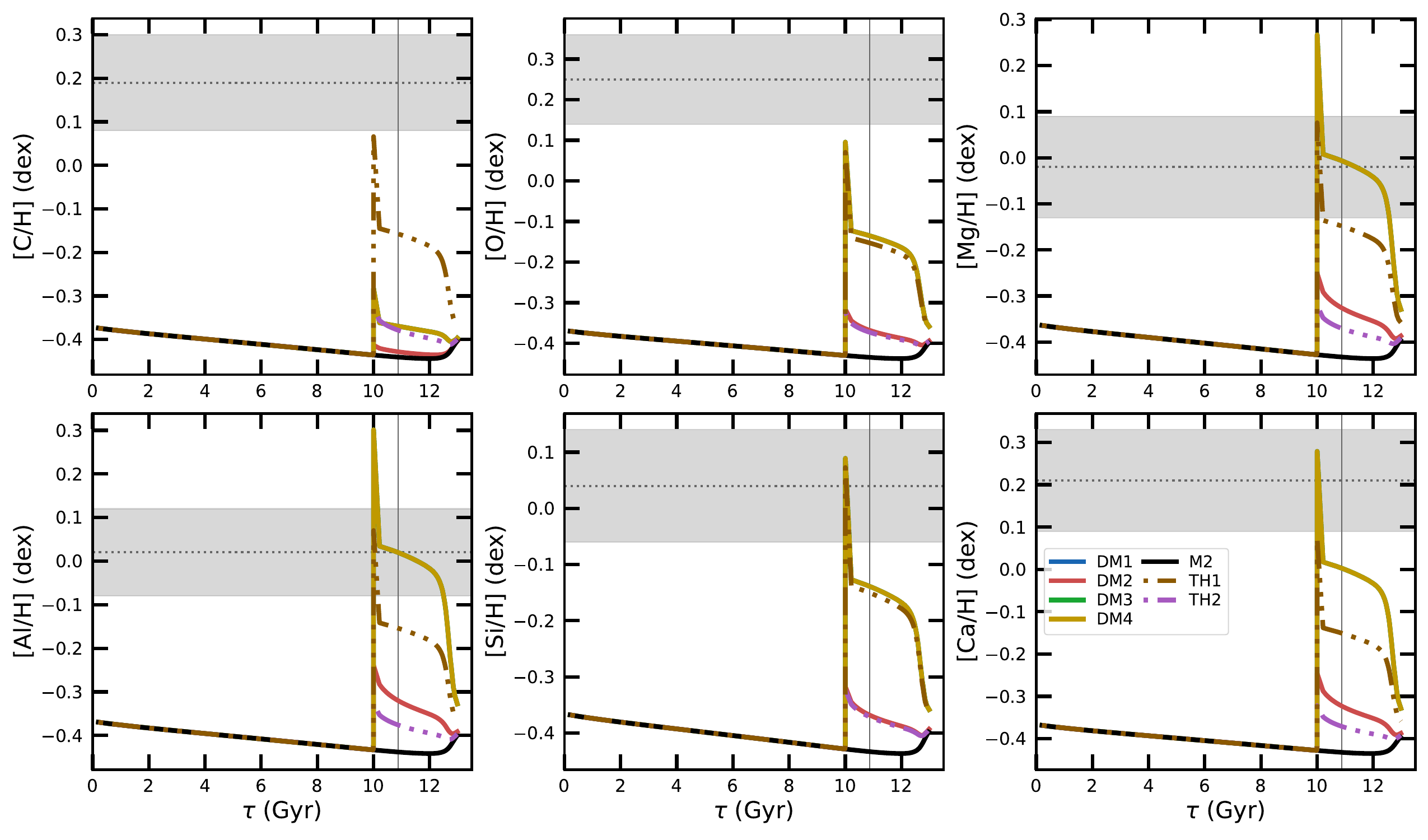}
    \caption{Comparison of surface elemental evolution for accretion events using  
    Cl-chondrite mixture (DM1, DM2, DM3, DM4 models of Table \ref{tab:acre_models_diff_mixture}). }
    \label{fig:accret_other_elements_diff_mix}
\end{figure*}

This CI-chondrite mixture has different elemental ratios, and their evolution is shown in Fig. \ref{fig:accret_other_elements_diff_mix}. O and Si exhibit similar abundances and evolution as in the  solar chemical mixture. 
In contrast,
 Mg, Al, and Ca, which have higher mass fractions in the CI-chondrite mixture, produce greater variations at the stellar surface. For C, the accretion event does not enrich the stellar surface as much as the accretion of the solar mixture does, indicating that its fraction in the CI-chondrite mixture is lower.

We also performed similar calculations using the chemical mixture presented by \citet{Wang2018}, representative of the bulk composition of Earth-like planets. While this mixture produces slight differences in iron and other elemental abundances, we notice that the behavior of lithium remains unchanged, leading to the same conclusions.

\section{Discussion}
\label{sec:discussion}
The chemical dichotomy of the HD~81809 system is extreme: however, beyond the accretion scenario, two other hypotheses must be considered. First, if both components form from the same environment, they are expected to have the same initial chemical composition. Studies by \cite{Deal2015,Ramirez2019,Liu2021} argued that chemical variations of 0.1-0.2~dex between stellar companions, can be explained by chemical transport mechanisms. For normal stars, expected to follow standard Galatic chemical evolution, chemical variations larger than 0.2~dex are not expected during stellar evolution. Standard mechanisms like atomic diffusion usually compete with other transport processes to produce only minor variations typically on order of 0.1-0.2 dex. In solar-type stars with $M<1.1~\Msun$, the convective envelope is deep enough to prevent extreme surface chemical discrepancies. Large chemical variations ($>0.5$ dex) are generally only possible in stars with very thin convective envelopes, such as F-type stars or more massive ones \citep{moedas2022}. For the stellar components of HD~818909 system this not expected to be created by atomic diffusion, unless any mechanisms to suppress the convective envelope and other chemical transport mechanisms.

The second possibility is that the two stars formed in different environments. HD~818909B could have formed in a more metal-rich environment than HD~818909A and was subsequently captured by the system. Following this scenario, the authors of Paper I inferred the properties of HD~818909B. However, they found setbacks: first, the star would be older than the universe, and the secondary star would be more massive than the primary component of the system, which is unrealistic. However, as the authors point out, these results need to be carefully analyzed, and it is possible to achieve a feasible result when taking into consideration the uncertainties.

In any case, an accretion event is a plausible scenario that could explain the observed chemical discrepancy. However, the high abundance of lithium detected and the possible presence of a debris disk may not be sufficient support. Other signals that could support an accretion event include the detection of abnormal surface rotation and magnetic activity. In the case of planetary accretion, angular momentum is transferred during the engulfment, which spins up the star \citep{Privitera2016}. This can consequently reactivate the magnetic dynamo \citep{Bellotti2026}. Unfortunately, there is currently no rotation or magnetic field detection for HD~818909B that can identify these signals.

Another possibility is that the star underwent a late 'pebble wave', which could have increased its metallicity. \cite{Kunitomo2021}  found that a pebble wave during the early stages of solar evolution could increase the metallicity of the Sun by 5\%. These changes allow the solar internal structure to be brought into closer alignment with heliosphere data. If this happened to our star during the pre-MS stage, it could prevent the significant enrichment of lithium due to the large convective envelope. However, due to this large envelope, the amount of material necessary to increase the abundances to the observed levels would be greater than that observed previously (more than double), as the material would be completely diluted \citep{Soares2025}. If this happened during the main sequence, the process would be the same as what we tested before. If the process is fast, it could be simulated as planet accretion; if the process is slow, it could resemble constant accretion.

\section{Conclusion}
\label{sec:conclusion}

The HD~81809 system is chemically peculiar, with a difference of 0.57 dex in $\FeH$ between its two stellar components. As mentioned in Paper I, the higher metallicity of HD~81809B could be evidence of an accretion event, supported by its high surface lithium value and the presence of a possible debris disk. This makes HD~81809B a unique target to test accretion in stellar models. 
We explored a range of accretion scenarios, considering different accretion masses and timings.

Our results indicate that it is possible to reproduce the observed surface $\mathrm{[Fe/H]}$ with a recent accretion event, occurring near the current stellar age, involving roughly 50~$\mathrm{M}\oplus$ of pure metals (with or without thermohaline convection). 
Although this is a large amount of metals for an accretion event, it is physically plausible, as some gas giants are expected to contain rocky cores up to $\sim 67~\mathrm{M}\oplus$ \citep{Sato2005,Wolf2007}.
An accretion event that reproduced the surface $\FeH$ could also help explain the lower effective temperature of HD~81809B.

However, to reproduce the observed values modeling accretion events occurring in the early stages of evolution, the required accreted mass exceeds the 250 $\Meart$, which is unlikely.  
Although we can reproduce the iron surface, we are not able to reproduce the observed lithium surface. The models predict that such a metal-rich accretion would over-enrich lithium at the surface; matching the observed lithium instead requires accreting less than 6~$\mathrm{M}_\oplus$. This tension highlights the need for precise knowledge of the accreted material’s chemical composition.

We verified that hydrogen and helium can be neglected during accretion events, as planetary accretion is insufficient to significantly alter their abundances in stellar envelopes. This justifies focusing exclusively on metal accretion.

To explain the lithium discrepancy, two hypotheses can be proposed. 
First, the observed lithium abundance is uncertain. Second, the chemical composition of the accreted material is
different from the stellar initial composition.

Tests using a CI-chondrite mixture \citep{Lodders2009} of our solar system, slightly increased both iron and lithium at the surface, but did not resolve the discrepancy.
The actual composition of the accreted material likely depends on the planet’s formation environment and may differ from the solar mixture.
It is important to note that we assumed the accreted material had the same chemical composition as that of our solar system, which may not be true. According to \cite{Fuhrmann2018} and Paper I, this system should have a higher abundance of alpha elements, suggesting that the solar mixture we adopted is incorrect. 

It is important to note that we used the solar chemical mixture from \cite{Asplund2009} in the stellar models. This could be a source of uncertainty, especially since the primary is an alpha-enhanced star. Using a different chemical mixture and an appropriate opacity table would affect the position of the stellar models in the HR diagram, which would affect the inferred stellar properties, such as mass and age. It may also affect the amount of material necessary to accrete in order to achieve the observed abundances of the secondary.

In summary, HD~81809B provides a valuable case for testing planetary accretion in stellar models. Our results confirm that metal accretion can explain the surface iron enhancement, but reproducing lithium requires further constraints on the composition of the engulfed material. Nevertheless, these are preliminary results that warrant further exploration. In Paper I, the authors used TESS data to obtain the global seismic parameter; however, they were unable to resolve individual frequencies. Re-observing the system for a longer duration could provide individual oscillation modes, especially for HD~81809A, which would give us stronger constraints on the stellar age of the system. The work of Paper I was also the first to provide spectral abundances for HD~81809B using model atmospheres computed with ATLAS9 \citep{kur93}. Detecting rotation and magnetic activity on HD~81809B would provide additional evidence of a planet being engulfed, however it would be challenging to disentangle these signals from those originating from the primary stellar component.

\begin{acknowledgements}
This work was funded by the European Union (ERC, MAGNIFY, Project 101126182 ). Views and opinions expressed are however those of the author(s) only and do not necessarily reflect those of the European Union or the European Research Council. Neither the European Union nor the granting authority can be held responsible for them.

\end{acknowledgements}

\bibliographystyle{aa}
\bibliography{references}

@article{Behmard2022,
    author = {Behmard, Aida and Sevilla, Jason and Fuller, Jim},
    title = {Planet engulfment signatures in twin stars},
    journal = {Monthly Notices of the Royal Astronomical Society},
    volume = {518},
    number = {4},
    pages = {5465-5474},
    year = {2022},
    month = {12},
}

@ARTICLE{Paxton2011,
       author = {{Paxton}, Bill and {Bildsten}, Lars and {Dotter}, Aaron and et al.},
        title = "{Modules for Experiments in Stellar Astrophysics (MESA)}",
      journal = {\apjs},
         year = 2011,
        month = jan,
       volume = {192},
       number = {1},
          eid = {3},
        pages = {3},
          doi = {10.1088/0067-0049/192/1/3},
archivePrefix = {arXiv},
       eprint = {1009.1622},
 primaryClass = {astro-ph.SR},
       adsurl = {https://ui.adsabs.harvard.edu/abs/2011ApJS..192....3P}
}

@ARTICLE{Asplund2009,
       author = {{Asplund}, Martin and {Grevesse}, Nicolas and {Sauval}, A. Jacques and {Scott}, Pat},
        title = "{The Chemical Composition of the Sun}",
      journal = {\araa},
         year = 2009,
        month = sep,
       volume = {47},
       number = {1},
        pages = {481-522},
          doi = {10.1146/annurev.astro.46.060407.145222},
archivePrefix = {arXiv},
       eprint = {0909.0948},
 primaryClass = {astro-ph.SR},
       adsurl = {https://ui.adsabs.harvard.edu/abs/2009ARA&A..47..481A}
}

@ARTICLE{Paxton2013,
       author = {{Paxton}, Bill and {Cantiello}, Matteo and {Arras}, Phil and {Bildsten}, Lars and {Brown}, Edward F. and {Dotter}, Aaron and {Mankovich}, Christopher and {Montgomery}, M.~H. and {Stello}, Dennis and {Timmes}, F.~X. and {Townsend}, Richard},
        title = "{Modules for Experiments in Stellar Astrophysics (MESA): Planets, Oscillations, Rotation, and Massive Stars}",
      journal = {\apjs},
         year = 2013,
        month = sep,
       volume = {208},
       number = {1},
          eid = {4},
        pages = {4},
          doi = {10.1088/0067-0049/208/1/4},
archivePrefix = {arXiv},
}

@ARTICLE{Paxton2015,
       author = {{Paxton}, Bill and {Marchant}, Pablo and {Schwab}, Josiah and {Bauer}, Evan B. and {Bildsten}, Lars and {Cantiello}, Matteo and {Dessart}, Luc and {Farmer}, R. and {Hu}, H. and {Langer}, N. and {Townsend}, R.~H.~D. and {Townsley}, Dean M. and {Timmes}, F.~X.},
        title = "{Modules for Experiments in Stellar Astrophysics (MESA): Binaries, Pulsations, and Explosions}",
      journal = {\apjs},
         year = 2015,
        month = sep,
       volume = {220},
       number = {1},
          eid = {15},
        pages = {15},
          doi = {10.1088/0067-0049/220/1/15},
archivePrefix = {arXiv},
       eprint = {1506.03146},
 primaryClass = {astro-ph.SR},
       adsurl = {https://ui.adsabs.harvard.edu/abs/2015ApJS..220...15P}
}

@ARTICLE{Paxton2018,
       author = {{Paxton}, Bill and {Schwab}, Josiah and {Bauer}, Evan B. and {Bildsten}, Lars and {Blinnikov}, Sergei and {Duffell}, Paul and {Farmer}, R. and {Goldberg}, Jared A. and {Marchant}, Pablo and {Sorokina}, Elena and {Thoul}, Anne and {Townsend}, Richard H.~D. and {Timmes}, F.~X.},
        title = "{Modules for Experiments in Stellar Astrophysics (MESA): Convective Boundaries, Element Diffusion, and Massive Star Explosions}",
      journal = {\apjs},
         year = 2018,
        month = feb,
       volume = {234},
       number = {2},
          eid = {34},
        pages = {34},
          doi = {10.3847/1538-4365/aaa5a8},
archivePrefix = {arXiv},
       eprint = {1710.08424},
 primaryClass = {astro-ph.SR},
       adsurl = {https://ui.adsabs.harvard.edu/abs/2018ApJS..234...34P}
}

@ARTICLE{Paxton2019,
       author = {{Paxton}, Bill and {Smolec}, R. and {Schwab}, Josiah and {Gautschy}, A. and {Bildsten}, Lars and {Cantiello}, Matteo and {Dotter}, Aaron and {Farmer}, R. and {Goldberg}, Jared A. and {Jermyn}, Adam S. and {Kanbur}, S.~M. and {Marchant}, Pablo and {Thoul}, Anne and {Townsend}, Richard H.~D. and {Wolf}, William M. and {Zhang}, Michael and {Timmes}, F.~X.},
        title = "{Modules for Experiments in Stellar Astrophysics (MESA): Pulsating Variable Stars, Rotation, Convective Boundaries, and Energy Conservation}",
      journal = {\apjs},
         year = 2019,
        month = jul,
       volume = {243},
       number = {1},
          eid = {10},
        pages = {10},
          doi = {10.3847/1538-4365/ab2241},
archivePrefix = {arXiv},
       eprint = {1903.01426},
 primaryClass = {astro-ph.SR},
       adsurl = {https://ui.adsabs.harvard.edu/abs/2019ApJS..243...10P}
}

@ARTICLE{Thoul,
       author = {{Thoul}, Anne A. and {Bahcall}, John N. and {Loeb}, Abraham},
        title = "{Element Diffusion in the Solar Interior}",
      journal = {\apj},
         year = 1994,
        month = feb,
       volume = {421},
        pages = {828},
          doi = {10.1086/173695},
archivePrefix = {arXiv},
       eprint = {astro-ph/9304005},
 primaryClass = {astro-ph},
       adsurl = {https://ui.adsabs.harvard.edu/abs/1994ApJ...421..828T}
}

@ARTICLE{Krishna1966,
       author = {{Krishna Swamy}, K.~S.},
        title = "{Profiles of Strong Lines in K-Dwarfs}",
      journal = {\apj},
         year = 1966,
        month = jul,
       volume = {145},
        pages = {174},
          doi = {10.1086/148752},
       adsurl = {https://ui.adsabs.harvard.edu/abs/1966ApJ...145..174K}
}

@ARTICLE{Iglesias1996,
       author = {{Iglesias}, Carlos A. and {Rogers}, Forrest J.},
        title = "{Updated Opal Opacities}",
      journal = {\apj},
         year = 1996,
        month = jun,
       volume = {464},
        pages = {943},
          doi = {10.1086/177381},
       adsurl = {https://ui.adsabs.harvard.edu/abs/1996ApJ...464..943I}
}

@ARTICLE{Rogers2002,
       author = {{Rogers}, F.~J. and {Nayfonov}, A.},
        title = "{Updated and Expanded OPAL Equation-of-State Tables: Implications for Helioseismology}",
      journal = {\apj},
         year = 2002,
        month = sep,
       volume = {576},
       number = {2},
        pages = {1064-1074},
          doi = {10.1086/341894},
       adsurl = {https://ui.adsabs.harvard.edu/abs/2002ApJ...576.1064R}
}

@BOOK{Cox1968,
       author = {{Cox}, J.~P. and {Giuli}, R.~T.},
        title = "{Principles of stellar structure}",
         year = 1968,
  publisher =  {Gordon \& Breach},      
       adsurl = {https://ui.adsabs.harvard.edu/abs/1968pss..book.....C}
}

@ARTICLE{moedas2022,
       author = {{Moedas}, Nuno and {Deal}, Morgan and {Bossini}, Diego and {Campilho}, Bernardo},
        title = "{Atomic diffusion and turbulent mixing in solar-like stars: Impact on the fundamental properties of FG-type stars}",
      journal = {\aap},
         year = 2022,
        month = oct,
       volume = {666},
          eid = {A43},
        pages = {A43},
          doi = {10.1051/0004-6361/202243210},
archivePrefix = {arXiv},
       eprint = {2207.02779},
 primaryClass = {astro-ph.SR},
       adsurl = {https://ui.adsabs.harvard.edu/abs/2022A&A...666A..43M}
}

@ARTICLE{Eggenberger2022,
       author = {{Eggenberger}, P. and {Buldgen}, G. and {Salmon}, S.~J.~A.~J. and {Noels}, A. and {Grevesse}, N. and {Asplund}, M.},
        title = "{The internal rotation of the Sun and its link to the solar Li and He surface abundances}",
      journal = {Nature Astronomy},
         year = 2022,
        month = may,
       volume = {6},
        pages = {788-795},
          doi = {10.1038/s41550-022-01677-0},
       adsurl = {https://ui.adsabs.harvard.edu/abs/2022NatAs...6..788E}
}

@ARTICLE{Ferguson2005,
       author = {{Ferguson}, Jason W. and {Alexander}, David R. and {Allard}, France and {Barman}, Travis and {Bodnarik}, Julia G. and {Hauschildt}, Peter H. and {Heffner-Wong}, Amanda and {Tamanai}, Akemi},
        title = "{Low-Temperature Opacities}",
      journal = {\apj},
         year = 2005,
        month = apr,
       volume = {623},
       number = {1},
        pages = {585-596},
          doi = {10.1086/428642},
archivePrefix = {arXiv},
       eprint = {astro-ph/0502045},
 primaryClass = {astro-ph},
       adsurl = {https://ui.adsabs.harvard.edu/abs/2005ApJ...623..585F}
}

@ARTICLE{Proffitt1991,
       author = {{Proffitt}, Charles R. and {Michaud}, Georges},
        title = "{Diffusion and Mixing of Lithium and Helium in Population II Dwarfs}",
      journal = {\apj},
         year = 1991,
        month = apr,
       volume = {371},
        pages = {584},
          doi = {10.1086/169923},
       adsurl = {https://ui.adsabs.harvard.edu/abs/1991ApJ...371..584P}
}

@ARTICLE{Theado2009,
       author = {{Th{\'e}ado}, S. and {Vauclair}, S. and {Alecian}, G. and {LeBlanc}, F.},
        title = "{Influence of Thermohaline Convection on Diffusion-Induced Iron Accumulation in a Stars}",
      journal = {\apj},
         year = 2009,
        month = oct,
       volume = {704},
       number = {2},
        pages = {1262-1273},
          doi = {10.1088/0004-637X/704/2/1262},
archivePrefix = {arXiv},
       eprint = {0908.1534},
 primaryClass = {astro-ph.SR},
       adsurl = {https://ui.adsabs.harvard.edu/abs/2009ApJ...704.1262T}
}

@ARTICLE{Deal2015,
       author = {{Deal}, Morgan and {Richard}, Olivier and {Vauclair}, Sylvie},
        title = "{Accretion of planetary matter and the lithium problem in the 16 Cygni stellar system}",
      journal = {\aap},
         year = 2015,
        month = dec,
       volume = {584},
          eid = {A105},
        pages = {A105},
          doi = {10.1051/0004-6361/201526917},
archivePrefix = {arXiv},
       eprint = {1509.06958},
 primaryClass = {astro-ph.SR},
       adsurl = {https://ui.adsabs.harvard.edu/abs/2015A&A...584A.105D}
}

@ARTICLE{Moedas2025,
       author = {{Moedas}, Nuno and {Deal}, Morgan and {Bossini}, Diego},
        title = "{Impact of radiative accelerations on the stellar characterization of FGK-type stars using spectroscopic and seismic constraints}",
      journal = {\aap},
         year = 2025,
        month = mar,
       volume = {695},
          eid = {A9},
        pages = {A9},
          doi = {10.1051/0004-6361/202453130},
archivePrefix = {arXiv},
       eprint = {2502.05025},
 primaryClass = {astro-ph.SR},
       adsurl = {https://ui.adsabs.harvard.edu/abs/2025A&A...695A...9M}
}

@ARTICLE{Townsley2004,
       author = {{Townsley}, Dean M. and {Bildsten}, Lars},
        title = "{Theoretical Modeling of the Thermal State of Accreting White Dwarfs Undergoing Classical Nova Cycles}",
      journal = {\apj},
         year = 2004,
        month = jan,
       volume = {600},
       number = {1},
        pages = {390-403},
          doi = {10.1086/379647},
archivePrefix = {arXiv},
       eprint = {astro-ph/0306080},
 primaryClass = {astro-ph},
       adsurl = {https://ui.adsabs.harvard.edu/abs/2004ApJ...600..390T}
}

@ARTICLE{Soares2025,
       author = {{Soares}, B.~M.~T.~B. and {Adibekyan}, V. and {Mordasini}, C. and {Deal}, M. and {Sousa}, S.~G. and {Delgado-Mena}, E. and {Santos}, N.~C. and {Dorn}, C.},
        title = "{Assessing the processes behind planet engulfment and its imprints}",
      journal = {\aap},
         year = 2025,
        month = jan,
       volume = {693},
          eid = {A47},
        pages = {A47},
          doi = {10.1051/0004-6361/202451399},
archivePrefix = {arXiv},
       eprint = {2411.13455},
 primaryClass = {astro-ph.EP},
       adsurl = {https://ui.adsabs.harvard.edu/abs/2025A&A...693A..47S}
}

@ARTICLE{Brown2013,
       author = {{Brown}, Justin M. and {Garaud}, Pascale and {Stellmach}, Stephan},
        title = "{Chemical Transport and Spontaneous Layer Formation in Fingering Convection in Astrophysics}",
      journal = {\apj},
         year = 2013,
        month = may,
       volume = {768},
       number = {1},
          eid = {34},
        pages = {34},
          doi = {10.1088/0004-637X/768/1/34},
archivePrefix = {arXiv},
       eprint = {1212.1688},
 primaryClass = {astro-ph.SR},
       adsurl = {https://ui.adsabs.harvard.edu/abs/2013ApJ...768...34B}
}

@ARTICLE{Mauro2026,
       author = {{Di Mauro}, Maria Pia and {Pezzotti}, Camilla and {Moedas}, Nuno and {Catanzaro}, Giovanni and {Maxted}, Pierre F.~L. and {Corsaro}, Enrico and {Reda}, Raffaele and {Scuflaire}, Richard and {Bonanno}, Alfio and {Giovannelli}, Luca and {Beck}, Paul G.},
        title = "{On the contradictory case of the binary system HD 81809 hosting two pulsating solar-like stars observed by TESS}",
      journal = {arXiv e-prints},
         year = 2026,
        month = jan,
          eid = {arXiv:2601.20652},
        pages = {arXiv:2601.20652},
archivePrefix = {arXiv},
       eprint = {2601.20652},
 primaryClass = {astro-ph.SR},
       adsurl = {https://ui.adsabs.harvard.edu/abs/2026arXiv260120652D}
}

@ARTICLE{Orlando2017,
       author = {{Orlando}, S. and {Favata}, F. and {Micela}, G. and {Sciortino}, S. and {Maggio}, A. and {Schmitt}, J.~H.~M.~M. and {Robrade}, J. and {Mittag}, M.},
        title = "{Fifteen years in the high-energy life of the solar-type star HD 81809. XMM-Newton observations of a stellar activity cycle}",
      journal = {\aap},
         year = 2017,
        month = sep,
       volume = {605},
          eid = {A19},
        pages = {A19},
          doi = {10.1051/0004-6361/201731301},
archivePrefix = {arXiv},
       eprint = {1707.06437},
 primaryClass = {astro-ph.SR},
       adsurl = {https://ui.adsabs.harvard.edu/abs/2017A&A...605A..19O}
}

@ARTICLE{Egeland2018,
       author = {{Egeland}, Ricky},
        title = "{Deconvolving the HD 81809 Binary: Rotational and Activity Evidence for a Subgiant with a Sun-like Cycle}",
      journal = {\apj},
         year = 2018,
        month = oct,
       volume = {866},
       number = {2},
          eid = {80},
        pages = {80},
          doi = {10.3847/1538-4357/aadf86},
archivePrefix = {arXiv},
       eprint = {1807.10870},
 primaryClass = {astro-ph.SR},
       adsurl = {https://ui.adsabs.harvard.edu/abs/2018ApJ...866...80E}
}

@ARTICLE{Duquennoy1988,
       author = {{Duquennoy}, A. and {Mayor}, M.},
        title = "{Duplicity in the solar neighbourhood. III. New spectroscopic elementsfor nine solar-type binary stars.}",
      journal = {\aap},
         year = 1988,
        month = apr,
       volume = {195},
        pages = {129-147},
       adsurl = {https://ui.adsabs.harvard.edu/abs/1988A&A...195..129D}
}

@ARTICLE{Fuhrmann2018,
       author = {{Fuhrmann}, Klaus and {Chini}, Rolf},
        title = "{Fossil Merger of a Population II Star}",
      journal = {\apj},
         year = 2018,
        month = may,
       volume = {858},
       number = {2},
          eid = {103},
        pages = {103},
          doi = {10.3847/1538-4357/aabaff},
       adsurl = {https://ui.adsabs.harvard.edu/abs/2018ApJ...858..103F}
}

@INPROCEEDINGS{Adibekyan2018,
       author = {{Adibekyan}, Vardan and {Sousa}, S{\'e}rgio G. and {Santos}, Nuno C.},
        title = "{Characterization of Exoplanet-Host Stars}",
    booktitle = {Asteroseismology and Exoplanets: Listening to the Stars and Searching for New Worlds},
         year = 2018,
       editor = {{Campante}, Tiago L. and {Santos}, Nuno C. and {Monteiro}, M{\'a}rio J.~P.~F.~G.},
       series = {Astrophysics and Space Science Proceedings},
       volume = {49},
        month = jan,
        pages = {225},
          doi = {10.1007/978-3-319-59315-9_12},
archivePrefix = {arXiv},
       eprint = {1711.01112},
 primaryClass = {astro-ph.EP},
       adsurl = {https://ui.adsabs.harvard.edu/abs/2018ASSP...49..225A}
}

@ARTICLE{Saffe2017,
       author = {{Saffe}, C. and {Jofr{\'e}}, E. and {Martioli}, E. and {Flores}, M. and {Petrucci}, R. and {Jaque Arancibia}, M.},
        title = "{Signatures of rocky planet engulfment in HAT-P-4. Implications for chemical tagging studies}",
      journal = {\aap},
         year = 2017,
        month = jul,
       volume = {604},
          eid = {L4},
        pages = {L4},
          doi = {10.1051/0004-6361/201731430},
archivePrefix = {arXiv},
       eprint = {1707.02180},
 primaryClass = {astro-ph.SR},
       adsurl = {https://ui.adsabs.harvard.edu/abs/2017A&A...604L...4S}
}

@ARTICLE{Oh2018,
       author = {{Oh}, Semyeong and {Price-Whelan}, Adrian M. and {Brewer}, John M. and {Hogg}, David W. and {Spergel}, David N. and {Myles}, Justin},
        title = "{Kronos and Krios: Evidence for Accretion of a Massive, Rocky Planetary System in a Comoving Pair of Solar-type Stars}",
      journal = {\apj},
         year = 2018,
        month = feb,
       volume = {854},
       number = {2},
          eid = {138},
        pages = {138},
          doi = {10.3847/1538-4357/aaab4d},
archivePrefix = {arXiv},
       eprint = {1709.05344},
 primaryClass = {astro-ph.SR},
       adsurl = {https://ui.adsabs.harvard.edu/abs/2018ApJ...854..138O}
}

@ARTICLE{Ramirez2019,
       author = {{Ram{\'\i}rez}, I. and {Khanal}, S. and {Lichon}, S.~J. and {Chanam{\'e}}, J. and {Endl}, M. and {Mel{\'e}ndez}, J. and {Lambert}, D.~L.},
        title = "{The chemical composition of HIP 34407/HIP 34426 and other twin-star comoving pairs}",
      journal = {\mnras},
         year = 2019,
        month = dec,
       volume = {490},
       number = {2},
        pages = {2448-2457},
          doi = {10.1093/mnras/stz2709},
archivePrefix = {arXiv},
       eprint = {1909.07460},
 primaryClass = {astro-ph.SR},
       adsurl = {https://ui.adsabs.harvard.edu/abs/2019MNRAS.490.2448R}
}

@ARTICLE{Nagar2020,
       author = {{Nagar}, Tushar and {Spina}, Lorenzo and {Karakas}, Amanda I.},
        title = "{The Chemical Signatures of Planetary Engulfment Events in Binary Systems}",
      journal = {\apjl},
         year = 2020,
        month = jan,
       volume = {888},
       number = {1},
          eid = {L9},
        pages = {L9},
          doi = {10.3847/2041-8213/ab5dc6},
       adsurl = {https://ui.adsabs.harvard.edu/abs/2020ApJ...888L...9N}
}

@ARTICLE{LiuF2014,
       author = {{Liu}, F. and {Asplund}, M. and {Ramirez}, I. and {Yong}, D. and {Melendez}, J.},
        title = "{A high-precision chemical abundance analysis of the HAT-P-1 stellar binary: constraints on planet formation.}",
      journal = {\mnras},
         year = 2014,
        month = jul,
       volume = {442},
        pages = {L51-L55},
          doi = {10.1093/mnrasl/slu055},
archivePrefix = {arXiv},
       eprint = {1404.2112},
 primaryClass = {astro-ph.SR},
       adsurl = {https://ui.adsabs.harvard.edu/abs/2014MNRAS.442L..51L}
}

@ARTICLE{Saffe2015,
       author = {{Saffe}, C. and {Flores}, M. and {Buccino}, A.},
        title = "{HD 80606: searching for the chemical signature of planet formation}",
      journal = {\aap},
         year = 2015,
        month = oct,
       volume = {582},
          eid = {A17},
        pages = {A17},
          doi = {10.1051/0004-6361/201526644},
archivePrefix = {arXiv},
       eprint = {1507.08125},
 primaryClass = {astro-ph.SR},
       adsurl = {https://ui.adsabs.harvard.edu/abs/2015A&A...582A..17S}
}

@ARTICLE{Mack2016,
       author = {{Mack}, III, Claude E. and {Stassun}, Keivan G. and {Schuler}, Simon C. and {Hebb}, Leslie and {Pepper}, Joshua A.},
        title = "{Detailed Abundances of Planet-hosting Wide Binaries. II. HD80606+HD80607}",
      journal = {\apj},
         year = 2016,
        month = feb,
       volume = {818},
       number = {1},
          eid = {54},
        pages = {54},
          doi = {10.3847/0004-637X/818/1/54},
archivePrefix = {arXiv},
       eprint = {1601.00018},
 primaryClass = {astro-ph.SR},
       adsurl = {https://ui.adsabs.harvard.edu/abs/2016ApJ...818...54M}
}

@ARTICLE{Liu2021,
       author = {{Liu}, Fan and {Bitsch}, Bertram and {Asplund}, Martin and {Liu}, Bei-Bei and {Murphy}, Michael T. and {Yong}, David and {Ting}, Yuan-Sen and {Feltzing}, Sofia},
        title = "{Detailed elemental abundances of binary stars: searching for signatures of planet formation and atomic diffusion}",
      journal = {\mnras},
         year = 2021,
        month = nov,
       volume = {508},
       number = {1},
        pages = {1227-1240},
          doi = {10.1093/mnras/stab2471},
archivePrefix = {arXiv},
       eprint = {2108.11001},
 primaryClass = {astro-ph.SR},
       adsurl = {https://ui.adsabs.harvard.edu/abs/2021MNRAS.508.1227L}
}

@ARTICLE{Cummings2017,
       author = {{Cummings}, Jeffrey D. and {Deliyannis}, Constantine P. and {Maderak}, Ryan M. and {Steinhauer}, Aaron},
        title = "{WIYN Open Cluster Study. LXXV. Testing the Metallicity Dependence of Stellar Lithium Depletion Using Hyades-aged Clusters. I. Hyades and Praesepe}",
      journal = {\aj},
         year = 2017,
        month = mar,
       volume = {153},
       number = {3},
          eid = {128},
        pages = {128},
          doi = {10.3847/1538-3881/aa5b86},
archivePrefix = {arXiv},
       eprint = {1702.03936},
 primaryClass = {astro-ph.SR},
       adsurl = {https://ui.adsabs.harvard.edu/abs/2017AJ....153..128C}
}

@ARTICLE{Sun2023,
       author = {{Sun}, Qinghui and {Deliyannis}, Constantine P. and {Steinhauer}, Aaron and {Anthony-Twarog}, Barbara J. and {Twarog}, Bruce A.},
        title = "{WIYN Open Cluster Study 89. M48 (NGC 2548) 2: Lithium Abundances in the 420 Myr Open Cluster M48 from Giants through K Dwarfs}",
      journal = {\apj},
         year = 2023,
        month = jul,
       volume = {952},
       number = {1},
          eid = {71},
        pages = {71},
          doi = {10.3847/1538-4357/acc5e3},
archivePrefix = {arXiv},
       eprint = {2303.09783},
 primaryClass = {astro-ph.SR},
       adsurl = {https://ui.adsabs.harvard.edu/abs/2023ApJ...952...71S}
}

@ARTICLE{Sevilla2022,
       author = {{Sevilla}, Cassie and {Behmard}, Aida and {Fuller}, Jim},
        title = "{Long-term lithium abundance signatures following planetary engulfment}",
      journal = {\mnras},
         year = 2022,
        month = nov,
       volume = {516},
       number = {3},
        pages = {3354-3365},
          doi = {10.1093/mnras/stac2436},
archivePrefix = {arXiv},
       eprint = {2207.13232},
 primaryClass = {astro-ph.SR},
       adsurl = {https://ui.adsabs.harvard.edu/abs/2022MNRAS.516.3354S}
}

@ARTICLE{Vauclair2012,
       author = {{Vauclair}, Sylvie and {Th{\'e}ado}, Sylvie},
        title = "{Thermohaline Instabilities inside Stars: A Synthetic Study Including External Turbulence and Radiative Levitation}",
      journal = {\apj},
         year = 2012,
        month = jul,
       volume = {753},
       number = {1},
          eid = {49},
        pages = {49},
          doi = {10.1088/0004-637X/753/1/49},
archivePrefix = {arXiv},
       eprint = {1205.1329},
 primaryClass = {astro-ph.SR},
       adsurl = {https://ui.adsabs.harvard.edu/abs/2012ApJ...753...49V}
}

@ARTICLE{Scuflaire2008,
       author = {{Scuflaire}, R. and {Th{\'e}ado}, S. and {Montalb{\'a}n}, J. and
         {Miglio}, A. and {Bourge}, P. -O. and {Godart}, M. and {Thoul}, A. and
         {Noels}, A.},
        title = "{CL{\'E}S, Code Li{\'e}geois d'{\'E}volution Stellaire}",
      journal = {\apss},
         year = 2008,
        month = aug,
       volume = {316},
       number = {1-4},
        pages = {83-91},
          doi = {10.1007/s10509-007-9650-1},
archivePrefix = {arXiv},
       eprint = {0712.3471},
 primaryClass = {astro-ph},
       adsurl = {https://ui.adsabs.harvard.edu/abs/2008Ap&SS.316...83S}
}

@PHDTHESIS{Donahue1993,
       author = {{Donahue}, Robert Andrew},
        title = "{Surface Differential Rotation in a Sample of Cool Dwarf Stars}",
       school = {New Mexico State University},
         year = 1993,
        month = jan,
       adsurl = {https://ui.adsabs.harvard.edu/abs/1993PhDT.........3D}
}

@ARTICLE{Donahue1996,
       author = {{Donahue}, Robert A. and {Saar}, Steven H. and {Baliunas}, Sallie L.},
        title = "{A Relationship between Mean Rotation Period in Lower Main-Sequence Stars and Its Observed Range}",
      journal = {\apj},
         year = 1996,
        month = jul,
       volume = {466},
        pages = {384},
          doi = {10.1086/177517},
       adsurl = {https://ui.adsabs.harvard.edu/abs/1996ApJ...466..384D}
}

@ARTICLE{Sato2005,
       author = {{Sato}, Bun'ei and {Fischer}, Debra A. and {Henry}, Gregory W. and {Laughlin}, Greg and {Butler}, R. Paul and {Marcy}, Geoffrey W. and {Vogt}, Steven S. and {Bodenheimer}, Peter and {Ida}, Shigeru and {Toyota}, Eri and {Wolf}, Aaron and {Valenti}, Jeff A. and {Boyd}, Louis J. and {Johnson}, John A. and {Wright}, Jason T. and {Ammons}, Mark and {Robinson}, Sarah and {Strader}, Jay and {McCarthy}, Chris and {Tah}, K.~L. and {Minniti}, Dante},
        title = "{The N2K Consortium. II. A Transiting Hot Saturn around HD 149026 with a Large Dense Core}",
      journal = {\apj},
         year = 2005,
        month = nov,
       volume = {633},
       number = {1},
        pages = {465-473},
          doi = {10.1086/449306},
archivePrefix = {arXiv},
       eprint = {astro-ph/0507009},
 primaryClass = {astro-ph},
       adsurl = {https://ui.adsabs.harvard.edu/abs/2005ApJ...633..465S}
}

@ARTICLE{Wolf2007,
       author = {{Wolf}, Aaron S. and {Laughlin}, Gregory and {Henry}, Gregory W. and {Fischer}, Debra A. and {Marcy}, Geoff and {Butler}, Paul and {Vogt}, Steve},
        title = "{A Determination of the Spin-Orbit Alignment of the Anomalously Dense Planet Orbiting HD 149026}",
      journal = {\apj},
         year = 2007,
        month = sep,
       volume = {667},
       number = {1},
        pages = {549-556},
          doi = {10.1086/503354},
       adsurl = {https://ui.adsabs.harvard.edu/abs/2007ApJ...667..549W}
}

@ARTICLE{Yildiz2024,
       author = {{Y{\i}ld{\i}z}, M. and {{\c{C}}elik Orhan}, Z. and {{\"O}rtel}, S. and {{\c{C}}ak{\i}r}, T.},
        title = "{Structure and composition of Jupiter, Saturn, Uranus, and Neptune under different constraints and distortion due to rotation}",
      journal = {\mnras},
         year = 2024,
        month = mar,
       volume = {528},
       number = {4},
        pages = {6881-6894},
          doi = {10.1093/mnras/stae476},
       adsurl = {https://ui.adsabs.harvard.edu/abs/2024MNRAS.528.6881Y}
}

@ARTICLE{Lodders2009,
       author = {{Lodders}, K. and {Palme}, H. and {Gail}, H.-P.},
        title = "{Abundances of the Elements in the Solar System}",
      journal = {Landolt B{\"o}rnstein},
         year = 2009,
        month = jan,
       volume = {4B},
        pages = {712},
          doi = {10.1007/978-3-540-88055-4_34},
archivePrefix = {arXiv},
       eprint = {0901.1149},
 primaryClass = {astro-ph.EP},
       adsurl = {https://ui.adsabs.harvard.edu/abs/2009LanB...4B..712L}
}

@ARTICLE{Wang2018,
       author = {{Wang}, Haiyang S. and {Lineweaver}, Charles H. and {Ireland}, Trevor R.},
        title = "{The elemental abundances (with uncertainties) of the most Earth-like planet}",
      journal = {\icarus},
         year = 2018,
        month = jan,
       volume = {299},
        pages = {460-474},
          doi = {10.1016/j.icarus.2017.08.024},
archivePrefix = {arXiv},
       eprint = {1708.08718},
 primaryClass = {astro-ph.EP},
       adsurl = {https://ui.adsabs.harvard.edu/abs/2018Icar..299..460W}
}

@inproceedings{kur93,
	adsurl = {https://ui.adsabs.harvard.edu/abs/1993ASPC...44...87K},
	author = {{Kurucz}, R.~L.},
	booktitle = {{IAU Colloq. 138: Peculiar versus Normal Phenomena in A-type and Related Stars}},
	editor = {{Dworetsky}, M.~M. and {Castelli}, F. and {Faraggiana}, R.},
	month = jan,
	pages = {87},
	series = {{Astronomical Society of the Pacific Conference Series}},
	title = {{A New Opacity-Sampling Model Atmosphere Program for Arbitrary Abundances}},
	volume = {44},
	year = 1993
}

@ARTICLE{Bellotti2026,
       author = {{Bellotti}, S. and {Pezzotti}, C. and {Buldgen}, G. and {Vidotto}, A.~A. and {Evensberget}, D. and {Magaudda}, E.},
        title = "{Eating planets makes you younger: The magnetic dynamo rejuvenation of GJ 504 by planetary engulfment}",
      journal = {\aap},
         year = 2026,
        month = mar,
       volume = {707},
          eid = {L12},
        pages = {L12},
          doi = {10.1051/0004-6361/202659144},
archivePrefix = {arXiv},
       eprint = {2602.22979},
 primaryClass = {astro-ph.SR},
       adsurl = {https://ui.adsabs.harvard.edu/abs/2026A&A...707L..12B}
}

@ARTICLE{Privitera2016,
       author = {{Privitera}, Giovanni and {Meynet}, Georges and {Eggenberger}, Patrick and {Vidotto}, Aline A. and {Villaver}, Eva and {Bianda}, Michele},
        title = "{Star-planet interactions. II. Is planet engulfment the origin of fast rotating red giants?}",
      journal = {\aap},
         year = 2016,
        month = oct,
       volume = {593},
          eid = {A128},
        pages = {A128},
          doi = {10.1051/0004-6361/201628758},
archivePrefix = {arXiv},
       eprint = {1606.08027},
 primaryClass = {astro-ph.EP},
       adsurl = {https://ui.adsabs.harvard.edu/abs/2016A&A...593A.128P}
}

@ARTICLE{Kunitomo2021,
       author = {{Kunitomo}, Masanobu and {Guillot}, Tristan},
        title = "{Imprint of planet formation in the deep interior of the Sun}",
      journal = {\aap},
         year = 2021,
        month = nov,
       volume = {655},
          eid = {A51},
        pages = {A51},
          doi = {10.1051/0004-6361/202141256},
archivePrefix = {arXiv},
       eprint = {2109.06492},
 primaryClass = {astro-ph.SR},
       adsurl = {https://ui.adsabs.harvard.edu/abs/2021A&A...655A..51K}
}

\begin{appendix}

\onecolumn

\end{appendix}

\end{document}